\documentclass[12pt]{article}
\usepackage{amsthm}
 \usepackage{caption}
 \usepackage{mathrsfs}
 \usepackage{footmisc}
 \usepackage{pstricks}
 \usepackage[colorlinks=true,citecolor=blue]{hyperref}
 \usepackage{enumerate}
\usepackage[shortlabels]{enumitem}
\usepackage[nohead]{geometry}
\usepackage{graphicx}
\usepackage{fourier}
\usepackage{microtype}
\usepackage{comment}
\usepackage{tikz}

\usetikzlibrary{trees}
\usepackage[authoryear]{natbib}
\usepackage{amsmath}
 
\usepackage{dirtytalk}
\usepackage{amsfonts}
\usetikzlibrary{shapes,backgrounds}
\usepackage{blindtext}
\usepackage{amssymb}
\theoremstyle{plain}

\newtheorem{theorem}{Theorem}
\newtheorem{definition}{Definition}
\newtheorem{proposition}{Proposition}
\newtheorem{lemma}{Lemma}
\newtheorem{corollary}{Corollary}
\theoremstyle{definition}

\usepackage{accents}

\newtheorem{contaclaim0}{Contaclaim0}
\newtheorem{contaclaim}{Contaclaim}
\newtheorem{contaclaim2}{Contaclaim2}
\newtheorem{contaclaim3}{Contaclaim3}

\theoremstyle{definition}

\usepackage{todonotes}

\tikzstyle{decision} = [rectangle, minimum height=18pt, minimum width=18pt,  fill=none, ultra thick, inner sep=0pt]
\tikzstyle{chance} = [circle, minimum width=18pt,  fill=none, ultra thick, inner sep=0pt]
\tikzstyle{line} = [draw=none]
\usetikzlibrary{calc}

\newtheorem{claim0}[contaclaim0]{Claim}
\newtheorem{claim}[contaclaim]{Claim}
\newtheorem{claim2}[contaclaim2]{Claim}
\newtheorem{claim3}[contaclaim3]{Claim}
\usepackage{etoolbox}
\AddToHook{env/proof/after}{}
\AddToHook{env/claim/after}{}

\newtheorem{inneraxiom}{Axiom}
\newtheorem*{inneraxiomnonumber}{\axiomname}
\providecommand{\axiomname}{}

\makeatletter
\NewDocumentEnvironment{asiom}{o}
 {%
  \IfNoValueTF{#1}
    {\inneraxiom}
    {
     \renewcommand{\axiomname}{#1}
     \def\@currentlabel{#1}
     \inneraxiomnonumber
    }%
 }
 {\IfNoValueTF{#1}{\endinneraxiom}{\endinneraxiomnonumber}}
\makeatother

\begin{document}
\title{Recursive Preferences and Ambiguity Attitudes\footnote{We thank the editor Faruk Gul and the anonymous referees whose thorough comments helped us to improve the quality of the paper. We also thank Lorenzo Bastianello, Arjada Bardhi, Antoine Bommier, Simone Cerreia-Vioglio, Paul Cheung, Joyee Deb, Tommaso Denti, Adam Dominiak, Francesco Fabbri, Takashi Hayashi, Chiaki Hara, Peter Klibanoff, Asen Kochov, Fran\c{c}ois Le Grand, Erik Madsen, Fabio Maccheroni, Alfonso Maselli, Hiroki Nishimura, Efe Ok, Luciano Pomatto, and Kota Saito for their comments. We thank participants at TUS 2022, RUD 2023, D-TEA 2023, SAET 2023, and Science of Decision Making 2023.
}}
\renewcommand{\thefootnote}{\fnsymbol{footnote}}
\author{Massimo Marinacci\footnote{Department of Decision Sciences and IGIER,  Università Bocconi, Via Roentgen, 1, Milan, Italy.  E-mail: massimo.marinacci@unibocconi.it} \and  Giulio Principi\footnote{Department of Economics, New York University, 269 Mercer St., 7th Floor, New York, NY 10003, USA. E-mail: gp2187@nyu.edu} \and Lorenzo Stanca\footnote{Department of ESOMAS and Collegio Carlo Alberto, Università di Torino, Corso Unione Sovietica, 218 Bis 10134, Turin, Italy. E-mail: lorenzomaria.stanca@unito.it}
}

\date{\today}

\maketitle
\begin{abstract}

Monotonicity and recursivity are central assumptions in intertemporal consumption problems under ambiguity. We show that monotone recursive preferences admit both a recursive and an ex-ante representation, and that the certainty equivalent functionals associated with these representations are translation invariant. Translation invariance implies that the decision maker’s ambiguity attitudes are constant, in the sense that they do not vary with the level of welfare or utility. Finally, we establish an equation linking the two certainty equivalent functionals that extends the notion of rectangularity for recursive multiple priors, which we call  \textit{generalized rectangularity}. This equation yields a simple, testable condition for dynamic consistency.
\end{abstract}



\section{Introduction}
\renewcommand*{\thefootnote}{\arabic{footnote}}

In many dynamic economic environments, decision makers face uncertainty that is hard to quantify. Governments set fiscal policy amid ambiguous growth and climate policy amid ambiguity about future climate conditions and their economic consequences. Households decide how much to consume and save without knowing the distribution of future income. Firms commit to long-lived investments while unsure  about the probability distribution of demand and technology shocks.  In such settings of \textit{ambiguity}, it is natural to ask what  principles should discipline preferences over uncertain intertemporal plans.

Monotonicity and recursivity are two central assumptions in intertemporal consumption problems under ambiguity. The alternatives the decision maker compares in these problems are uncertain streams of outcomes, which we call \textit{consumption plans}. Monotonicity requires that, state by state, consumption plans delivering higher discounted utility are preferred. Under recursivity, preferences satisfy dynamic consistency, history independence, and stationarity, which permit the use of dynamic programming methods. Dynamic consistency means choices made ex-ante are not later overturned; history independence means that the ranking of continuation plans is unaffected by the common consumption history preceding them; and stationarity means that these rankings are invariant to a common shift in time.

In our main result, we show that preferences that are both monotone and recursive admit two equivalent representations: an ex‑ante one and a recursive one. Moreover, the certainty equivalents (i.e., monotone functionals) associated with both representations are translation invariant. Translation invariance means that adding a constant to every outcome raises the certainty equivalent by the same constant; behaviorally, it implies constant absolute ambiguity aversion, i.e., the ambiguity aversion of the agent is not affected by (absolute) changes in wealth or utility. We further show that the two certainty equivalents are linked by a law of iterated nonlinear expectations---an equation that generalizes rectangularity for the recursive  multiple‑prior model (see \citeauthor{epstein2003iid}, \citeyear{epstein2003iid}; \citeauthor{epstein2003recursive}, \citeyear{epstein2003recursive}). We refer to this equation as \textit{generalized rectangularity}.

We present a simple environment in which a shock  determines the level of consumption each period. The decision maker has preferences over consumption plans, that is, sequences of state-contingent consumption levels that specify consumption for every possible realization of the sequence of shocks. Ambiguity in consumption arises from the decision maker not knowing the probability distribution of these shocks. Initially, for simplicity, we assume that beliefs about shocks remain constant over time---the assumption of independently and indistinguishably distributed (IID) ambiguity (see \citeauthor{epstein2003iid}, \citeyear{epstein2003iid}). In simple terms, it is as if 
the decision maker observes a single draw from a new Ellsberg urn every period. Nevertheless, as we show later in the paper, this assumption can be fully relaxed.

Theorem \ref{monrecu} shows that monotone recursive preferences admit two representations—an ex-ante one and a recursive one. Under the ex-ante representation, the decision maker compares consumption plans by aggregating the induced state-contingent discounted utilities via an \textit{ex-ante certainty equivalent}, whose properties reflect the decision maker’s ignorance about the true probability distribution. In mathematical terms, preferences $\succsim$ admit the following representation: for all consumption plans $h=(h_t)_{t\geq 0}$ and $g=(g_t)_{t\geq 0}$,
$$
h\succsim g 
\;\Longleftrightarrow\;
I_0\!\left(\sum_{t\ge 0}\beta^{t}\,u(h_t)\right)
\;\ge\;
I_0\!\left(\sum_{t\ge 0}\beta^{t}\,u(g_t)\right),
$$
where $I_0$ is the (ex-ante) certainty equivalent (a monotone function), $u$ is the (per-period) utility function, and $\beta\in(0,1)$ is the discount factor. The ex-ante certainty equivalent $I_0$ captures both beliefs and ambiguity attitudes about the full sequence of shocks. 

By contrast, the recursive representation evaluates a consumption plan through an iterative process. In each period, the decision maker combines the current consumption value with a discounted certainty equivalent of future utility. This certainty equivalent, referred to as the \textit{one-step-ahead certainty equivalent}, captures the decision maker's beliefs and attitudes toward ambiguity regarding the next period shock. Formally, preferences $\succsim$ of the decision maker admit the following representation
$$
h\succsim g \Longleftrightarrow V(h)=u(h_0)+\beta I_{+1}(V\circ h^1)\geq V(g)=u(g_0)+\beta I_{+1}(V\circ g^1),
$$
where $V$ is the value of the consumption plan, $h^1,g^1$ are the consumption plans shifted ahead by one period, and $I_{+1}$ is the  one-step-ahead certainty equivalent (again, a monotone function).

Theorem \ref{monrecu} shows that, under monotonicity and recursivity, both certainty equivalents $I_0$ and $I_{+1}$ must be \textit{translation invariant}. This property means that if a deterministic amount of wealth or utility is added to every possible outcome of an ambiguous prospect, the certainty equivalent increases by that same amount. 

Furthermore, Theorem \ref{monrecu} also establishes a link between the two representations. The intuition is that dynamic consistency---implied by recursivity---requires that ex-ante optimal decisions cannot be overturned. In our setting, this requires the choices induced by $I_0$ and $I_{+1}$ to be consistent. This consistency is ensured by the generalized rectangularity condition we derive
\begin{equation}\label{eqintro:gr}
I_0\left(\sum_{t\geq 1}\beta^t u(h_t)\right)=\beta I_{+1}\left(s\mapsto \frac{1}{\beta}I_{0}\left(\sum_{t\geq 1}\beta^t u(h^s_t)\right)\right)
\end{equation}
for all consumption plans $h$, where $h^s$ denotes the consumption plan conditional on the realization of shock $s$. To gain intuition, suppose for a moment that both $I_0$ and $I_{+1}$ are standard expectations. Then \eqref{eqintro:gr} reduces to the law of iterated expectations, which is precisely the dynamic consistency requirement under subjective expected utility. More generally, \eqref{eqintro:gr} can thus be thought of as a {law of iterated nonlinear expectations}.

We then examine the implications of Theorem \ref{monrecu}. Translation invariance of the ex-ante certainty equivalent has important implications for the decision maker's attitudes toward ambiguity.   Corollary \ref{monrecucor}  shows that  translation invariance   restricts the decision maker’s preferences to exhibit \textit{constant absolute ambiguity aversion}. In other words, the decision maker's ambiguity aversion remains unchanged with respect to absolute changes in wealth or utility levels.  

We also discuss the implications of our result when preferences, in addition, feature \emph{uncertainty aversion}. Here,  uncertainty aversion refers to convexity of preferences, or equivalently, to a preference for hedging (\citeauthor{gilboa1989maxmin}, \citeyear{gilboa1989maxmin}).  Corollary \ref{variationalcor} shows, that because of translation invariance of the certainty equivalents, the only monotone recursive preferences satisfying  uncertainty aversion  are the dynamic formulation of variational preferences \citep{maccheroni2006dynamic}. Variational preferences are a widely studied class of preferences under ambiguity that subsumes, among others, multiplier preferences introduced by \cite{hansen2001robust}  and have been axiomatically founded by \cite{maccheroni2006ambiguity}. Our result points to a nontrivial modeling restriction in applied work: if a researcher commits to monotone and recursive preferences that exhibit uncertainty aversion, then she cannot use any model other than variational preferences.

In  Section~\ref{sec:genrect} we  examine another important implication of Theorem \ref{monrecu}. Specifically, we show that generalized rectangularity yields a simple, testable condition for dynamic consistency. We can think of generalized rectangularity as a functional equation: treat the ex-ante certainty equivalent $I_0$ as given, and regard the one-step-ahead certainty equivalent $I_{+1}$ as the unknown to be solved for. Proposition~\ref{lem:genrectyieldsrecu} shows that, if a decision maker’s preferences admit the ex-ante representation induced by a translation invariant $I_0$ and the generalized rectangularity equation \eqref{eqintro:gr} has a solution $I_{+1}$, then those preferences are dynamically consistent. We then illustrate how Theorem~\ref{monrecu} and Proposition~\ref{lem:genrectyieldsrecu}, taken together, can be used to design an experimental test of dynamic consistency.

Finally, we show how our results extend to non-IID settings that can accommodate learning. There are many environments in which agents learn about the ambiguity they face. For example, governments today know more about climate change than in the past.  Proposition \ref{monrecutgeneral2} extends Theorem \ref{monrecu} to this more general setting. In this case, preferences can change  as new information arrives, so that both certainty equivalents---one-step-ahead and ex-ante---vary with the realized sequence of shocks. We show that each certainty equivalent remains translation invariant. Generalized rectangularity becomes a set of conditions, one for each possible shock sequence, instead of a single equation.

\vspace{0.15cm}

\noindent

\textbf{Related literature}. Dynamic choice under ambiguity is central in many economic settings.\footnote{See \cite{ilut2022modeling} for a review.} For instance, \cite{ju2012ambiguity} model ambiguity about an unknown parameter governing the dynamics of consumption growth; \cite{millner2013scientific} study ambiguity about climate sensitivity, which determines how global temperatures respond to higher carbon emissions; and \cite{ilut2014ambiguous} consider ambiguity about future total-factor productivity.

A large literature develops the theoretical foundations of recursive preferences under ambiguity; to keep the review succinct, we focus on the contributions most closely related to our analysis. To state our main result, we begin by imposing a \emph{constant-beliefs} assumption, often referred to as IID ambiguity.\footnote{Later in the paper we show that our results extend beyond the IID framework.} This IID ambiguity framework was introduced by \cite{epstein2003iid} in the context of recursive multiple-priors preferences. \cite{strzalecki2013temporal} adopts the same setting to study recursive ambiguity preferences and the timing of uncertainty resolution. Particularly closely related is \cite{bommier2017monotone}, who investigate the role of monotonicity for recursive preferences under both risk and ambiguity. They establish translation invariance of the (one-step-ahead) certainty equivalent in the recursive representation, but do not analyze the ex-ante representation. Section~\ref{sec:CAAA} discusses the connection with their results in more detail.

    The ex-ante representation we study is also analyzed in \cite{bommier2019ambiguity} and \cite{kochov2015time}. Both papers work in a dynamic ambiguity environment with an arbitrary information structure, and, more generally than us, do not rely on objectively risky lotteries. \cite{bommier2019ambiguity} study the interaction between ambiguity and endogenous discounting under a notion of stationarity (``path stationarity'') that differs from ours.\footnote{See Section~5.1 in \cite{bommier2019ambiguity} for a detailed comparison.} Using axioms tailored to that setting, they obtain an ex-ante representation in which the associated certainty equivalent is translation invariant and satisfies a homotheticity restriction tied to the endogenous discount factor (see their Theorem 1). In a similar setting, \cite{kochov2015time} provides an axiomatization of  intertemporal maxmin expected utility and highlights how weakening the invariance to timing can accommodate richer ambiguity attitudes, including variational-type specifications. Theorem~2 in \cite{kochov2015time} is conceptually close to our Corollary~\ref{variationalcor}; we compare the two results in the discussion immediately following Corollary~\ref{variationalcor}. Overall, both \cite{bommier2019ambiguity} and \cite{kochov2015time} primarily focus on the ex-ante representation. Our analysis complements theirs by studying the ex-ante and recursive representations simultaneously, and by connecting them through the generalized rectangularity condition.

Our results on generalized rectangularity contribute to identifying preference conditions that ensure dynamic consistency under ambiguity. Closest are \cite{epstein2003iid,epstein2003recursive}, who introduce \emph{rectangularity} in multiple-priors models as a restriction on the set of priors that guarantees dynamic consistency. Relatedly, \cite{maccheroni2006dynamic} derive dynamic consistency requirements within the variational framework, while \cite{chandrasekher2022dual} (see their Online Appendix) provide analogous conditions for invariant biseparable preferences. Finally, \cite{savochkin2022dynamic} study recursive smooth ambiguity preferences: they show that recursivity forces constant absolute ambiguity aversion and, under this restriction, characterize dynamic consistency through a condition on the decision maker’s beliefs. Our contribution complements these approaches by extending the rectangularity logic beyond multiple-priors and by expressing the relevant restriction directly as a functional relation between the certainty equivalents of the ex-ante and recursive representations.

\section{Mathematical notation and framework}\label{Notation}
Let $K\subseteq \mathbb{R}$, $Y$ be a nonempty set, and $\mathcal{A}$ an algebra on $Y$. We denote by $B(K,Y,\mathcal{A})$ the set of bounded $\mathcal{A}$-measurable functions taking values in $K$ which we endow with the supnorm $\lVert \cdot \rVert_{\infty}$. For simplicity, we set $B(Y,\mathcal{A}):=B(\mathbb{R},Y,\mathcal{A})$. We identify constant functions with constants. Let $\mathbf{B}$ be a subset of $B(K,Y,\mathcal{A})$ such that $K\subseteq \mathbf{B}$. A functional $I:\mathbf{B}\to \mathbb{R}$ is said to be \textit{normalized} if $I(k)=k$ for all $k\in K$. We say that $I$ is \textit{monotone} if $I(\xi)\geq I(\varphi)$ whenever $\xi\geq \varphi$, for all $\xi,\varphi\in \mathbf{B}$. If $I$ is monotone and normalized, then it is said to be a \textit{certainty equivalent}. Moreover, we say that $I$ is \textit{translation invariant} if $I(\xi+k)=I(\xi)+k$, for all $\xi\in \mathbf{B}$ and  $k\in \mathbb{R}$ such that $\xi+k\in \mathbf{B}$. 

We denote by $\bigtriangleup(Y)$ the set of finitely additive probability measures on $Y$. If $\mathcal{A}$ is a $\sigma$-algebra, we denote by $\bigtriangleup^\sigma(Y)$ the set of countably additive probability measures on $Y$. For all nonempty sets $A$, we will denote by $A^{\infty}:=\prod_{t\geq 1}A$ its countably infinite Cartesian product and by $2^A$ its power set.

\subsection{Framework}

In the first part of the paper, we adopt a framework analogous to \cite{strzalecki2013temporal}, but with an infinite horizon (see also Section A.1 of \citeauthor{bommier2017monotone}, \citeyear{bommier2017monotone}). Specifically, we work in the stationary IID ambiguity environment introduced by \cite{epstein2003iid}. Under IID ambiguity, the uncertainty the decision maker faces in period $t$ is the same as the uncertainty faced in period $t+1$; the only relevant distinction is the timing at which uncertainty is resolved. This structure rules out inference: it is as if, in each period, the decision maker observes a single draw from a new Ellsberg urn. We adopt this setting to streamline the analysis in the first part of the paper and to present the main result without the additional notation required for general recursive preferences. Nevertheless, the results extend beyond the IID case, as we formally establish in Section \ref{sec:relax}.

Let $S$ be a  nonempty finite set representing the states of the world (\textit{shocks}) that realize in each period. The associated algebra of events is $\Sigma:=2^S$. The full state space, $\Omega:=S^{\infty}$, consists of states $\omega \in \Omega$ that specify a complete history $\left(s_1, s_2, \ldots\right)$. In each period $t\geq 1$, the individual knows the partial history $s^t:=\left(s_1, \ldots, s_t\right)$. Accordingly, information evolves as described by the filtration $\left(\mathcal{G}_t\right)_{t\geq 0}$ on $\Omega$ where $\mathcal{G}_0:=\{\emptyset, \Omega\}$ and, for all $t\geq 1$,
$$
\mathcal{G}_t:=\left\lbrace A\times S^{\infty}:A\subseteq S^t \right\rbrace.
$$
Equivalently, $\mathcal{G}_t$ is the sigma-algebra generated by the cylinder events $A_1\times \cdots \times A_t\times S^\infty$ with each $A_i\subseteq S$. Finally, let $\mathcal{G}:=\sigma\left(\bigcup_{t\geq 0} \mathcal{G}_t\right)$, i.e., the smallest sigma-algebra generated by the union of the sigma-algebras of the filtration $\left(\mathcal{G}_t\right)_{t\geq 0}$. 
\par\medskip
Let $C$ denote the set of one-period consumption outcomes. A (deterministic) consumption level is an element $c\in C$. We assume that $C$  is a compact metrizable space with at least two distinct elements. The set of lotteries over $C$ is denoted by $X=\bigtriangleup^{\sigma}(C)$, endowed with the weak convergence topology and the related Borel sigma-algebra. As customary, we identify $C$ as a subset of $X$. A \textit{consumption plan} is an $X$-valued, $\left(\mathcal{G}_t\right)_{t\geq 0}$-adapted stochastic process, that is, a sequence $h=\left(h_t\right)_{t\geq 0}$ such that each $h_t: \Omega \rightarrow X$ is $\mathcal{G}_t$-measurable. The set of all consumption plans is denoted by $\mathbf{H}$ and  is endowed with the product topology (i.e., topology of pointwise convergence). We denote by $\mathbf{D}:=X^{\infty}$ the set of all deterministic consumption plans and identify $X$ as a subset of $\mathbf{D}$ where each $x\in X$ is seen as the constant consumption plan that yields the lottery $x$ in each period. Given a continuous function $u:X\rightarrow \mathbb{R}$ and $\beta\in (0,1)$, $U:\mathbf{D}\rightarrow\mathbb{R}$ denotes the discounted utility function, i.e., the map defined by  $$U:d\mapsto\sum_{t\geq 0} \beta^t u(d_t).$$
Analogously, we define 
$$
U(h):\omega\mapsto \sum_{t\geq 0}\beta^t u(h_t(\omega))
$$
for all $h\in \mathbf{H}$ and $U(\mathbf{H}):=\left\lbrace U(h):h\in \mathbf{H}\right\rbrace$.

For all $h \in \mathbf{H}$ and $s \in S$, we define the \textit{conditional consumption plan} $h^s \in \mathbf{H}$ by
$$
h^s\left(s_1, s_2,s_3, \ldots\right)=h\left(s, s_2,s_3, \ldots\right)=\left(h_0, h_1\left(s, s_2, \ldots\right),h_2(s,s_2,s_3,\ldots), \ldots\right)
$$ 
for all $(s_1,s_2,s_3,\ldots)\in \Omega$. In words, given $h\in \mathbf{H}$ and a state $s\in S$, the conditional consumption plan $h^s$ is the consumption plan obtained from $h$ when the decision maker knows that in the first period $s$ realizes. Conditional consumption plans allow us to define the continuation of consumption plans. Given $h=(h_0,h_1,h_2,\ldots)\in \mathbf{H}$ and $s\in S$, the \textit{continuation of} $h$, denoted by $h^{s,1}$, is defined as
$$
h^{s,1}\left(s_1, s_2, \ldots\right)=\left(h_1\left(s, s_2, \ldots\right), h_2\left(s, s_2, \ldots\right), \ldots\right)
$$
for all $\left(s_1, s_2, \ldots\right) \in \Omega$. The continuation $h^{s,1}$ is the consumption plan shifting $h$ forward by one period and knowing that in the first period state $s\in S$ realizes. Figure \ref{fig:continuation} illustrates an example of a continuation of a plan $h$ in which uncertainty resolves after one period. For all $h\in \mathbf{H}$, we will denote by $h^1$ the mapping $s\mapsto h^{s,1}$. 
\begin{figure}
    \centering
      \begin{tikzpicture}[grow=right,->,scale=0.85,font=\footnotesize,baseline={(current bounding box.center)}]
        \tikzstyle{level 1}=[level distance=25mm, sibling distance=25mm]
        \tikzstyle{level 2}=[level distance=25mm, sibling distance=12mm]
        \tikzstyle{level 3}=[level distance=18mm]
        \tikzstyle{solid node}=[circle,draw,inner sep=1.5,fill=black]
        \node(0)[solid node,label=above:{$h_0$}]{}
        child{node[solid node,label=below left:{$h_1(T)$}]{} 
            child{node[solid node,label=below:{$h_2(T)$}]{} 
            }
            edge from parent node[left,xshift=4,yshift=-5]{$T$}
        }
        child{node[solid node,label=above left:{$h_1(H)$}]{}
            child{node[solid node,label=above:{$h_2(H)$}]{} 
            }
            edge from parent node[left,xshift=4,yshift=5]{$H$}
        };
    \end{tikzpicture}
    \hspace{0.75cm}
 \begin{tikzpicture}[grow=right,->,scale=0.85,font=\footnotesize,baseline={(current bounding box.center)}]
        \tikzstyle{level 1}=[level distance=25mm, sibling distance=25mm]
        \tikzstyle{level 2}=[level distance=25mm, sibling distance=12mm]
        \tikzstyle{level 3}=[level distance=18mm]
        \tikzstyle{solid node}=[circle,draw,inner sep=1.5,fill=black]
        \node(0)[solid node,label=above:{$h_1(H)$}]{}
        child{node[solid node,label=above left:{$h_2(H)$}]{}
            child{node[solid node,label=above:{$h_3(H)$}]{} edge from parent node[left,shift={(0,-0.2)}]{}}
        };
    \end{tikzpicture}
    \caption{Continuation $h^{H,1}$ for $S=\{H,T\}$ when uncertainty resolves after one period.}
    \label{fig:continuation}
\end{figure}

Finally, we define the concatenation operation. For all lotteries $x$ and consumption plans $h$, the concatenation $\left(x,h\right)$ is defined as
\begin{equation}\label{concatenation}
(x,h)(s_1,s_2,\ldots)=\left(x,h(s_2,s_3,\ldots)\right)
\end{equation}
for all $(s_1,s_2,\ldots)\in \Omega$. Therefore, concatenation is obtained by inserting an additional consumption level in the initial period and delaying the resolution of uncertainty by one period.  This definition differs from that adopted by \cite{kochov2015time} and \cite{bommier2019ambiguity},\footnote{In particular, \cite{kochov2015time} and \cite{bommier2019ambiguity} define concatenation as $(x,h)(s_1,s_2,\ldots)=(x,h(s_1,s_2,\ldots))$.} and this distinction is important for separating our results from theirs. The notion of concatenated act employed here coincides with that used in \cite{bommier2017monotone}.\footnote{See their discussion on pages 1455--1456 for a detailed comparison of the alternative notions of concatenation used here and in \cite{bommier2019ambiguity} and \cite{kochov2015time}.} A simple yet important observation is that for all $x\in X$, $h\in \mathbf{H}$, and $s\in S$, $(x,h)^{s,1}=h$, i.e., the continuation of the concatenation $(x,h)$ is exactly $h$. 


\subsection{Preferences}
\label{intconsumptionchoice}
We consider a decision maker whose preferences are described by a binary relation $\succsim$ over $\mathbf{H}$. A function $V:\mathbf{H}\rightarrow\mathbb{R}$ \textit{ represents } $\succsim$ if $$h\succsim g\iff V(h)\geq V(g)$$ for all $h,g\in\mathbf{H}$. 

Our starting point is recursive preferences over consumption plans with an additively separable representation characterized by three components: a certainty equivalent $I_{+1}$, capturing ambiguous beliefs and ambiguity attitudes; a utility function $u$, representing preferences over consumption; and a discount factor $\beta$, reflecting time preferences. We call $I_{+1}$ the \textit{one-step-ahead certainty equivalent}. As we explained earlier, initially we assume no learning, meaning $I_{+1}$ remains unchanged as new information arrives. In Section~\ref{sec:relax}, we generalize our results to accommodate learning.

\begin{definition}\label{recursive_definition}
A binary relation $\succsim$ admits an \textnormal{IID separable recursive representation} if there exists a tuple $(V,I_{+1},u,\beta)$ such that $V:\mathbf{H}\to\mathbb{R}$ represents $\succsim$ and
\begin{equation}\label{recursivebommier}
V(h)=u(h_0)+\beta I_{+1}\left(V \circ h^1\right)
\end{equation}
where, $u:X\rightarrow\mathbb{R}$ is an affine, continuous, and nonconstant function, $\beta\in(0,1)$, and $I_{+1}:B(U(\mathbf{D}),S,\Sigma)\rightarrow\mathbb{R}$ is a certainty equivalent.
\end{definition}
Throughout the text we adopt the normalization $u(X)=[\beta-1,1-\beta]$.\footnote{By affinity $u$ is cardinally unique and by compactness and continuity $u(X)$ is bounded. Therefore, the normalization is without loss of generality. It implies that throughout $U(\mathbf{D})=[-1,1]$.} These preferences correspond to three main substantial axioms: dynamic consistency, history independence, and stationarity \cite[Section~3.3]{epstein1992behavior}.  The assumption of IID ambiguity  is captured  by the next axiom. It imposes a ``constant-beliefs'' condition, combining time consistency with a form of state independence.\footnote{The standard notion of dynamic consistency requires that  if in addition $h^s\succ g^s$ for some $s\in S$, then we have $h\succ g$ (see, for example,  \citeauthor{de2022static}, \citeyear{de2022static}, axiom D3). We consider the weaker notion, as it poses no complications. Our results also hold with the stronger notion with the addition that the one-step-ahead certainty equivalent would be strictly monotone.}
\begin{asiom}[State independent dynamic consistency (SIDC)]\label{axi:stateinddyncons} For all $h,g\in \mathbf{H}$ with $h_0=g_0$,
$$
\left[\forall s \in S, h^s \succsim g^s\right] \Longrightarrow h \succsim g.
$$
\end{asiom}
\noindent Stationarity expresses Koopmans' idea that \say{the passage of time does not have an effect on preferences.}   

\begin{asiom}[Stationarity (S)]\label{axi:stationarity} For all $x\in X$ and $h, g \in \mathbf{H}$, $h\succsim g$ if and only if $(x,h)\succsim (x,g)$.
\end{asiom}
Notice that our notion of stationarity requires not only invariance toward postponing consumption, but also delaying the resolution of uncertainty. Indeed, the consumption plan  $(x,h)$, as defined in equation \eqref{concatenation}, is obtained by adding one initial period of consumption and postponing the timing of resolution of uncertainty by one period, so that consumption at $t=2$  depends only on the information revealed in that same period.  Therefore, our stationarity axiom differs from the one considered in \cite{bommier2019ambiguity} and \cite{kochov2015time}, because of the distinct notion of concatenation we employ here.

History independence means that preferences do not depend on past consumption. In our setting, history independence is implied by time separability combined with the other axioms. 
\begin{asiom}[Time separability (TS)]\label{axi:timesep}
For all $x,y,x',y'\in X$ and $d, d'\in \mathbf{D}$, $(x,y,d)\sim (x',y',d)$ if and only if $(x,y,d')\sim (x',y',d')$.
\end{asiom}
 The remaining axioms  are basic requirements which guarantee the existence of a nontrivial and continuous representation. 
\begin{asiom}[Weak order (WO)]\label{axi:weakorder}
$\succsim$ is complete and transitive.
\end{asiom}
\begin{asiom}[Continuity (C)]\label{axi:cont} For all $g\in \mathbf{H}$, the sets
$$
\left\lbrace h\in \mathbf{H}:h\succsim g \right\rbrace\ \textnormal{and}\ \left\lbrace h\in \mathbf{H}: g\succsim h\right\rbrace
$$
are closed with respect to the product topology.
\end{asiom} 

\begin{asiom}[Nontriviality (NT)]\label{axi:nontriv} There exist $x,y\in X$ such that $x\succ y$.
\end{asiom} 
We study recursive preferences within a dynamic version of the Anscombe–Aumann framework. Consequently, we also impose an independence axiom restricted to  deterministic consumption plans.
\begin{asiom}[Independence for deterministic prospects (IDP)]\label{axi:indepdetprosp} For all $d,d',d''\in \mathbf{D}$ and $\alpha\in (0,1)$,
$$
d\sim d'\Longrightarrow \alpha d+(1-\alpha)d''\sim \alpha d'+(1-\alpha)d''.
$$
\end{asiom}

\section{Monotonicity and recursive preferences}\label{main_results}
Monotonicity is a basic rationality tenet of decision making under ambiguity (\citeauthor{cerreia2011rational}, \citeyear{cerreia2011rational}). Here we  consider the  standard intertemporal version  of monotonicity  typically adopted in the literature on dynamic choice (see, e.g., \citealp{epstein2003recursive}, \citealp{maccheroni2006dynamic}, and \citealp{bastianello2022choquet}).   
\begin{asiom}[Monotonicity (M)]\label{axi:mon}For all $h,g\in\mathbf{H}$,
$$
\left[\forall\omega\in \Omega,\ \left(h_t(\omega)\right)_{t\geq 0}\succsim \left(g_t(\omega)\right)_{t\geq 0}\right]\implies h\succsim g.
$$
\end{asiom}
When preferences over deterministic consumption streams are represented by exponentially discounted utility, monotone preferences admit an \textit{ex-ante} representation. That is, there exists an (ex-ante) certainty equivalent $I_0$ such that
$$
h \mapsto I_0\left(\sum_{t \geq 0} \beta^{t} u(h_t)\right)
$$
represents $\succsim$. Our main result characterizes IID separable recursive preferences that also admit an ex-ante representation. It shows how monotonicity and recursivity interact across the two representations, and how the certainty equivalent functionals at the core of each representation are linked.

\begin{theorem}\label{monrecu}
Let $\succsim$ be a binary relation on $\mathbf{H}$. The following are equivalent:
\begin{enumerate}[label=(\roman*)]
\item   $\succsim$ satisfies axioms \hyperref[axi:weakorder]{WO}, \hyperref[axi:cont]{C}, \hyperref[axi:nontriv]{NT}, \hyperref[axi:indepdetprosp]{IDP}, \hyperref[axi:stateinddyncons]{SIDC}, \hyperref[axi:stationarity]{S}, \hyperref[axi:timesep]{TS}, and \hyperref[axi:mon]{M}.
\item  $\succsim$  admits an IID separable recursive representation   $(V,I_{+1},u,\beta)$ with $I_{+1}$ translation invariant, $V$ continuous in the product topology, and there exists a translation invariant  certainty equivalent $I_0:U(\mathbf{H})\rightarrow\mathbb{R}$  such that $\succsim$ is represented by 
$$h\mapsto I_0\left(\sum_{t\geq 0} \beta^t u(h_t)\right).$$ 
\end{enumerate}
Moreover, $(I_0,I_{+1})$ satisfies
\begin{equation}
\label{generalrectangularity}
I_0\left(\sum_{t\geq 1}\beta^t u(h_t)\right)=\beta I_{+1}\left(s\mapsto \frac{1}{\beta}I_{0}\left(\sum_{t\geq 1}\beta^t u(h^s_t)\right)\right)
\end{equation}
for all $h\in \mathbf{H}$.
\end{theorem}
The proof proceeds in three steps. First, we use monotonicity (axiom~\hyperref[axi:mon]{M}) to construct the \emph{ex-ante} certainty equivalent $I_{0}$. Second, we show that the axioms of recursive preferences---particularly  stationarity  (axiom~\hyperref[axi:stationarity]{S})---imply translation invariance of $ I_0 $. Using this result and the recursive representation of preferences, we then establish that $ I_0 $ and $ I_{+1} $ are linked through equation~\eqref{generalrectangularity}, \textit{generalized rectangularity}. Finally, using the derived generalized rectangularity, we conclude that $ I_{+1} $ must also satisfy translation invariance.

We now examine in more depth the interpretation and the consequences of this result. First, we discuss the behavioral implications of translation invariance of the certainty equivalent $I_0$. Then, we discuss the relation between generalized rectangularity and dynamic consistency. Finally, we provide an extension of Theorem \ref{monrecu} that allows for dynamic updating of preferences.

\subsection{Translation invariance and ambiguity attitudes}\label{sec:CAAA}
We explore the main implications of  translation invariance for ambiguity attitudes.
\vspace{-15pt}
\subsubsection*{Constant absolute ambiguity aversion}
\vspace{-7pt}
First we discuss the notion of \textit{constant absolute ambiguity aversion} as introduced by \cite{grant2013mean}.
\begin{definition}\label{CAAA_def}
A binary relation $\succsim$ on $\mathbf{H}$ exhibits \textnormal{constant absolute ambiguity aversion} if for all $h\in \mathbf{H}$, $x,y,z\in X$, and $\alpha \in(0,1)$, $$\alpha h+(1-\alpha) x \succsim \alpha y+(1-\alpha) x \Longrightarrow \alpha h+(1-\alpha) z \succsim \alpha y+(1-\alpha) z.$$
\end{definition}
\noindent In words, constant absolute ambiguity aversion requires that whenever an uncertain alternative is preferred to a sure outcome, \say{adding} the same certain alternative to both does not invert the preference. In the static setting, absolute ambiguity attitudes have been thoroughly studied  by \cite{xue2020} and \cite{fabbri2024absolute}, and by \cite{cerreia2019ambiguity} in terms of utility and wealth changes, respectively.

\begin{sloppypar} When a binary relation is represented by a certainty equivalent functional and expected utility over lotteries, constant absolute ambiguity aversion is characterized by the translation invariance of the certainty equivalent. Therefore, Theorem \ref{monrecu}, by ensuring the translation invariance of the ex-ante certainty equivalent $I_0$, yields a first relevant restriction.\end{sloppypar}
\begin{corollary}\label{monrecucor}
If\hspace{0.1cm} $\succsim$ satisfies axioms \hyperref[axi:weakorder]{WO}, \hyperref[axi:cont]{C}, \hyperref[axi:nontriv]{NT}, \hyperref[axi:indepdetprosp]{IDP}, \hyperref[axi:stateinddyncons]{SIDC}, \hyperref[axi:stationarity]{S}, \hyperref[axi:timesep]{TS}, and \hyperref[axi:mon]{M}, then $\succsim$ exhibits constant absolute ambiguity aversion.
\end{corollary}
Therefore, Theorem \ref{monrecu} and Corollary \ref{monrecucor} highlight an important modeling trade-off. If one wants to allow for changing absolute ambiguity aversion under the tractability of recursivity, then monotonicity should be weakened. 

A subtle aspect is that the ambiguity attitudes of the decision maker are characterized by the properties of the ex-ante certainty equivalent $I_0$. In particular, any consumption plan $h=(h_0,h_1,\ldots)$ can be seen as an \textit{act} $h:\Omega\to \mathbf{D}$, the usual alternatives of static decision problems under ambiguity. Hence, the certainty equivalent $I_0$ represents \textit{static preferences} over these acts. We can therefore talk about ambiguity attitudes using the lifetime utility of the agent as the utility over deterministic acts. 

In contrast, a priori knowledge of the certainty equivalent $I_{+1}$ is insufficient to describe a  decision maker's uncertainty attitudes.  Indeed, $I_{+1}$ only contains information about attitudes on consumption plans that resolve all the uncertainty at $t=1$ and yield a constant stream of consumption thereafter.  

\vspace{0.15cm}
\noindent \textbf{Relationship with the literature}. \cite{bommier2017monotone} consider a more general formulation of recursive preferences. Specifically, they examine general nonseparable recursive preferences under the additional assumption of monotonicity, both in settings of objective uncertainty with temporal lotteries and in settings of subjective uncertainty. Notably, in the case of subjective uncertainty, they do not rely on the Anscombe--Aumann-type assumptions that we employ. They demonstrate that monotonicity is equivalent to the translation invariance of the one-step-ahead certainty equivalent.

Although \cite{bommier2017monotone} work in a more general framework, our analysis complements theirs by delivering additional results in our setting.  First, we introduce an  {ex-ante} certainty equivalent, $I_{0}$, and show that it must be translation invariant. Furthermore, this invariance {implies} constant absolute ambiguity aversion (CAAA), whereas the translation invariance of $I_{+1}$ alone does not. In addition, we propose a new dynamic consistency condition---generalized rectangularity---that connects the ex-ante and one-step-ahead representations. 

\vspace{-15pt}
\subsubsection*{Uncertainty averse preferences}
\vspace{-7pt}
We next examine the implications of Theorem~\ref{monrecu} under the additional requirement of \emph{uncertainty aversion}, meaning  convexity of preferences. In other words, objective mixtures of consumption plans are preferred to the original ones (see, for example, \citeauthor{cerreia2011uncertainty}, \citeyear{cerreia2011uncertainty}). The major implication is that  under monotonicity, the only recursive preferences that satisfy uncertainty aversion belong to the variational class of \cite{maccheroni2006ambiguity}.  

Before providing the results we need some more notation.
 We say that $c:\bigtriangleup(\Omega)\to \left[0,\infty\right]$ is \textit{grounded} if its infimum value is zero. In addition, we say that $c$ is a \textit{cost function} if it is convex, grounded, and (weakly) lower semicontinuous.
\begin{corollary}\label{variationalcor}
Suppose that $\succsim$  satisfies axioms \hyperref[axi:weakorder]{WO}, \hyperref[axi:cont]{C}, \hyperref[axi:nontriv]{NT}, \hyperref[axi:indepdetprosp]{IDP}, \hyperref[axi:stateinddyncons]{SIDC}, \hyperref[axi:stationarity]{S}, \hyperref[axi:timesep]{TS},  \hyperref[axi:mon]{M}, and
\begin{equation}\label{uncertaintyaversion}
h \sim g \Rightarrow \alpha h+(1-\alpha) g \succsim h,
\end{equation}
for all $h, g \in \mathbf{H}$ and $\alpha \in(0,1)$.
Then there exist cost functions $c_{+1}:\bigtriangleup(S)\rightarrow \left[0,\infty\right]$ and $c_0:\bigtriangleup(\Omega)\rightarrow \left[0,\infty\right]$ such that $\succsim$ admits as recursive representation
$$V:h\mapsto u(h_0)+\beta \min_{\ell\in\bigtriangleup(S)}\left\lbrace\mathbb{E}_{\ell} \left[V \circ h^1\right]+c_{+1}(\ell)\right\rbrace
$$ 
and as ex-ante representation
$$h\mapsto \min_{P\in\bigtriangleup(\Omega)}\left\lbrace \mathbb{E}_P\left[\sum_{t\geq 0}\beta^{t}u(h_t)\right]+c_0(P)\right\rbrace.$$
\end{corollary}
Condition \eqref{uncertaintyaversion} corresponds to the well-known  axiom of  uncertainty aversion  from \cite{gilboa1989maxmin}. In this specification, the decision maker evaluates a consumption plan using expected utility under the least favorable probabilistic model, while penalizing models according to their plausibility, thereby capturing concerns about misspecification (see \citeauthor{maccheroni2006dynamic}, \citeyear{maccheroni2006ambiguity}).

We note that this result is distinct from Theorem 2 in \cite{kochov2015time}. That result derives translation invariance from a combination of unbounded utility and history independence. Together with intertemporal hedging---which corresponds to the quasiconcavity of the ex-ante certainty equivalent---these assumptions yield the same ex-ante variational representation that we obtain. Our result, however, does not rely on unbounded utility. Instead, it relies on monotonicity along with all the axioms that characterize recursive preferences to establish translation invariance of the certainty equivalent.

\subsection{Generalized rectangularity: a simple test  of dynamic consistency}\label{sec:genrect}
Here we discuss another important consequence of Theorem \ref{monrecu} regarding equation \eqref{generalrectangularity}, which links the ex-ante and one-step-ahead certainty equivalents. In particular, generalized rectangularity provides a \emph{necessary and sufficient} condition for dynamic consistency within the class of monotone preferences considered here. This fact follows by Theorem \ref{monrecu} and the next result.
\begin{proposition}\label{lem:genrectyieldsrecu}
Suppose $I_0:U(\mathbf{H})\to\mathbb{R}$ is a translation invariant certainty equivalent and that there exists $I_{+1}:B\!\left(U(\mathbf{D}),S,\Sigma\right)\to\mathbb{R}$ such that
$$
I_0\left(\sum_{t\geq 1}\beta^t u(h_t)\right)=\beta I_{+1}\left(s\mapsto \frac{1}{\beta}I_{0}\left(\sum_{t\geq 1}\beta^t u(h^s_t)\right)\right)
$$
for all $h\in \mathbf{H}$. If  $\succsim$ is a binary relation such that
$$
h\succsim g\Longleftrightarrow I_0\left(\sum_{t\geq 0}\beta^tu(h_t)\right)\geq I_0\left(\sum_{t\geq 0}\beta^tu(g_t)\right)
$$
for all $h,g\in \mathbf{H}$, then $\succsim$ satisfies \hyperref[axi:stateinddyncons]{SIDC}.
\end{proposition}
\noindent Proposition \ref{lem:genrectyieldsrecu} yields a simple sufficient condition for dynamic consistency.   If preferences admit an ex-ante representation with $I_0$ translation invariant, and if there exists a  function $I_{+1}$ such that the pair $(I_0,I_{+1})$ satisfies generalized rectangularity, then   preferences satisfy (state-independent) dynamic consistency.

Thus, Theorem \ref{monrecu} and Proposition \ref{lem:genrectyieldsrecu} together contribute to the literature on experimental tests of dynamic consistency (for a review, see, e.g., \citeauthor{Sprenger2015}, \citeyear{Sprenger2015}) by delivering a simple testable restriction of dynamic consistency. This observation yields a two-step empirical procedure. First, using choice data, the analyst elicits the ex-ante representation---namely $u$, $\beta$, and the certainty equivalent $I_0$. If the ex-ante representation admits a solution to the generalized rectangularity equation, then $\succsim$ satisfies axiom~\hyperref[axi:stateinddyncons]{SIDC}.

 \subsection{Relaxing  IID Ambiguity}\label{sec:relax}
In the preceding sections, we have characterized recursive preferences under the assumption of IID ambiguity, where the decision maker’s beliefs remain constant over time. However,   many practical decision problems involve learning---beliefs change as new information arrives. 

In this section, we extend our analysis by relaxing the IID assumption, introducing a framework in which recursive preferences evolve through learning. We show that the core of our Theorem \ref{monrecu} remains unchanged. In particular, even in this more general framework, we still retrieve the implications of monotonicity and recursivity in terms of translation invariance of the ex-ante and one-step-ahead certainty equivalents, as well as the generalized rectangularity condition.

To do so, we need additional notation. Let 
$$
H = \{s_0\}\cup \bigcup_{t=1}^{\infty} S^t
$$
be the set of all histories, where each $s^t\in H $ is a history of length $ t\geq 0 $ and $s_0$ denotes the initial history. For all histories 
$s^t =(s_1, s_2, \dots, s_t), s^{t'}=(s'_1, s'_2, \dots, s'_{t'}) \in H$  the concatenated history $(s^t, s^{t'}) \in H$
is defined as
$$
(s^t, s^{t'}) = (s_1, s_2, \dots, s_t, s'_1, s'_2, \dots, s'_{t'}),
$$
To allow for learning, we let preferences adapt to the realized history of shocks.  Formally, consider a family of binary relations $(\succsim_{s^t})_{s^t\in H}$ satisfying the following:
\begin{asiom}[Conditional preference (CP)]\label{axi:condpref}
   For all $s^t\in H$ and $h,g \in \mathbf{H}$, if $h_\tau(s^t,\cdot)=g_\tau(s^t,\cdot)$ for all $\tau\geq t$, then $h\sim_{s^t} g$. Moreover, for all  $s^t\in H,$ $\mathbf{x}\in X^t$, and $d,d'\in\mathbf{D}$ $$d\succsim_{s^0} d'\iff (\mathbf{x},d)\succsim_{s^t} (\mathbf{x},d').$$
\end{asiom}

\noindent The first condition asserts that once history $s^t$ has been realized, only the continuation of consumption beyond $t$ matters—any differences off the path of $s^t$ are irrelevant.  The second point requires that  tastes are time‑invariant.

For $s^t\in H$, $x\in X$, and $h\in \mathbf{H}$, we define $(x,h)_{s^t}$ as
$$
((x,h)_{s^t})_t(s^t)=x\ \textnormal{and}\ ((x,h)_{s^t})_{t+m}(s^t,s'_{t+1},s'_{t+2},\ldots)=h_{t+m-1}(s^t,s'_{t+2},\ldots)
$$
for all $m\geq 1$ and $(s^t,s'_{t+1},s'_{t+2},\ldots)\in \Omega$. Since we will always assume \hyperref[axi:condpref]{CP}, subject to measurability, $(x,h)_{s^t}$ can be defined arbitrarily before period $t$ and for any $(r_1,r_2,\ldots)\in \Omega$ with $r^t\neq s^t$. By \hyperref[axi:condpref]{CP}, these choices do not affect the ranking expressed by $\succsim_{s^t}$.

For all $t\geq 0$, the function $U_t:\mathbf{D}\to \mathbb{R}$ is defined by:
$$
U_t(d)=\sum_{\tau\geq t}\beta^{\tau-t}u(d_\tau).
$$
Moreover, for all $s^t\in H$, we define
$$
U_t(h|s^t):(s'_1,s'_2,\ldots)\mapsto \sum_{\tau\geq t}\beta^{\tau-t}u(h_\tau(s^t,s'_{t+1},\ldots))
$$
and $U_t(\mathbf{H}|s^t):=\left\lbrace U_t(h|s^t):h\in \mathbf{H} \right\rbrace$. In this setting, we consider the following general formulation of time separable recursive preferences.
 \begin{definition}\label{recursive_definition_modified}
A collection of binary relations $(\succsim_{s^t})_{s^t\in H}$ admits a \textnormal{separable recursive representation} if there exists a collection $\left(V_{s^{t}},I_{+1,s^{t}},u,\beta\right)_{s^{t}\in H}$ such that each $V_{s^t}:\mathbf{H}\to\mathbb{R}$ represents $\succsim_{s^t}$,
\begin{equation}\label{generalrecursiverep}
V_{s^t}\left( h\right) = u\left(h_t\left(s^t\right)\right) + \beta I_{+1,s^t}\left(V_{(s^t,\cdot)}\left(h\right)\right),
\end{equation}
and $V_{s^t}((x,h)_{s^t})=u(x)+\beta V_{s^t}(h)$, where $u:X\to\mathbb{R}$ is an affine, continuous, and nonconstant function, $\beta\in(0,1)$, and $\bigl(I_{+1,s^t}\bigr)_{s^t\in H}$ is a collection of functions, where each $I_{+1,s^t}:B(U_t(\mathbf{D}),S,\Sigma)\to\mathbb{R}$ is a certainty equivalent.
\end{definition}
The IID case in Definition \ref{recursive_definition} corresponds to $I_{+1,s^t}=I_{+1}$ for all $s^t\in H$, i.e., the one-step-ahead certainty equivalent does not depend on  the realized sequence of shocks. The main difference with respect to the IID formulation is a weakening of \hyperref[axi:stateinddyncons]{state independent dynamic consistency} to a more general form of dynamic consistency.\footnote{Note that we use the following slight abuse of notation: for a history $s^t$, $h_t(s^t)$ denotes the common value of $h_t(\omega)$ over all paths $s^\infty=(s_1,s_2,\ldots)\in S^\infty$ whose first $t$ components coincide with $s^t$.}
  \begin{asiom}[Dynamic consistency (DC)]\label{axi:dynacons}
 For all $s^t\in H, h,g\in \mathbf{H}$ with $h_t(s^t)=g_t(s^t)$ 
$$
\left[\forall s \in S, h \succsim_{(s^t,s)} g\right] \Longrightarrow h \succsim_{s^t} g.
$$
\end{asiom}

In this setting, we are able to obtain the following generalization of Theorem \ref{monrecu}.
\begin{proposition}\label{monrecutgeneral2}
Let $(\succsim_{s^t})_{s^t\in H}$ be a  collection of binary relations on $\mathbf{H}$. The following are equivalent:
\begin{enumerate}[label=(\roman*)]
\item for all $s^t\in H$,    $\succsim_{s^t}$ satisfies axioms \hyperref[axi:weakorder]{WO}, \hyperref[axi:cont]{C}, \hyperref[axi:nontriv]{NT}, \hyperref[axi:indepdetprosp]{IDP}, \hyperref[axi:stationarity]{S}, \hyperref[axi:timesep]{TS},  \hyperref[axi:mon]{M} and $(\succsim_{s^t})_{s^t\in H}$ satisfies \hyperref[axi:condpref]{CP} and \hyperref[axi:dynacons]{DC}.
\item  $(\succsim_{s^t})_{s^t\in H}$ admits a  separable recursive representation  $\left(V_{s^{t}},I_{+1,s^{t}},u,\beta\right)_{s^{t}\in H}$  with each  $I_{+1,s^t}$ translation invariant  and there  exists     a translation invariant  certainty equivalent  $I_{s^t}:U_t(\mathbf{H}|s^t)\rightarrow\mathbb{R}$ such that $\succsim_{s^t}$ is represented by 
$$h\mapsto I_{s^t}\left(\sum_{\tau\geq t} \beta^{\tau-t} u(h_\tau(s^t,\cdot))\right).$$ 
\end{enumerate}
Moreover, each  $(I_{s^t},I_{+1,s^t})$ satisfies
\begin{equation}\;\label{eq:generalrect}
I_{s^t}\left(\sum_{m\geq 1} \beta^{m} u(h_{t+m}(s^t,\cdot))\right)=\beta I_{+1,s^t}\left(s\mapsto\frac{1}{\beta}I_{ (s^{t},s)}\left(\sum_{m\geq 1} \beta^{m} u(h_{t+m}(s^t,s,\ldots))\right)\right)
\end{equation}
for all $h\in \mathbf{H}$.
\end{proposition}
Therefore, this result shows that  each preference $\succsim_{s^t}$ will also satisfy constant absolute ambiguity aversion, independently of the information received.  The main difference with respect to Theorem \ref{monrecu} is  generalized rectangularity, equation \eqref{eq:generalrect}, which is now more complex. Indeed,  once learning is allowed,  generalized rectangularity becomes a collection of equations that depend on the realized history. 
\section{Concluding remarks}
Monotonicity and recursivity are natural principles in intertemporal decision-making. We have shown that, together, they yield a tractable representation of dynamic preferences and facilitate their elicitation. In particular, they imply constant absolute ambiguity aversion, a tractable class of ambiguity attitudes. If, in addition, preferences are uncertainty averse in the sense of a preference for hedging, then these assumptions restrict them to the recursive variational class.

Future research may explore specific instances of our representation. Generalized rectangularity could also be employed to derive versions of the law of large numbers and the central limit theorem for specific models of ambiguous preferences, as shown by \cite{epstein2003iid} for recursive multiple priors.

\section*{Appendix}

We start with a simple lemma that will be employed in the later sections.

\begin{lemma}\label{extension}
Let $Y$ be a nonempty set, and $\mathcal{A}$ and algebra on $Y$. Suppose $K\subseteq \mathbb{R}$ is an interval with $0\in K$ and $\mathbf{B}\subseteq B(Y,\mathcal{A})$ is convex with $K\subseteq \mathbf{B}$. If $I:\mathbf{B}\to \mathbb{R}$ is monotone, normalized, and
\begin{equation}\label{eq_inequality1}
\forall \xi,\varphi\in \mathbf{B},\ \forall a,b\in \mathbb{R},\ \xi+a\geq \varphi+b \Longrightarrow I(\xi)+a\geq I(\varphi)+b,
\end{equation}
then there exists a monotone, normalized, and translation invariant functional $\bar{I}:B(Y,\mathcal{A})\to \mathbb{R}$ such that $\bar{I}_{|\mathbf{B}}=I$. Moreover, if $I$ is concave, then $\bar{I}$ can be taken concave.
\end{lemma}
\begin{proof}
Define $\bar{I}:B(Y,\mathcal{A})\to [-\infty,\infty]$ as follows:
$$
\bar{I}(\xi)=\sup\left\lbrace I(\varphi)+k:\varphi\in \mathbf{B}, k\in \mathbb{R},\ \varphi+k\leq \xi \right\rbrace
$$
for all $\xi\in B(Y,\mathcal{A})$. We first observe that $\bar{I}$ is an extension of $I$. To see this, let $\xi\in \mathbf{B}$ and $\varphi\in \mathbf{B}$, $k\in \mathbb{R}$ be such that $\xi\geq \varphi+k$. By \eqref{eq_inequality1} we have that $I(\xi)\geq I(\varphi)+k$, and hence by the arbitrariness of $\varphi$ and $k$ we have $I(\xi)\geq \bar{I}(\xi)$. The converse inequality, $\bar{I}(\xi)\geq I(\xi)$, is trivial as $\xi\in \mathbf{B}$. Now, we show that $\bar{I}$ is finite. To see this first notice that the set $$\left\lbrace I(\varphi)+k:\varphi\in \mathbf{B}, k\in \mathbb{R},\ \varphi+k\leq \xi \right\rbrace$$ is nonempty. Indeed, if $\xi\in B(Y,\mathcal{A})$, then $\xi$ is bounded from below. Thus, let $k\in K$, there must exist $a\in \mathbb{R}$ such that $k+a\leq \xi$. Thus $\bar{I}>-\infty$. Now let $\xi\in B(Y,\mathcal{A})$ and $\varphi\in \mathbf{B}$, $k\in \mathbb{R}$ such that $\varphi+k\leq  \sup_{y\in Y} \xi(y)<\infty$. Then, since $0\in \mathbf{B}$ and $I$ is normalized, by \eqref{eq_inequality1}, we have
$$
I(\varphi)+k\leq I(0)+\sup_{y\in Y} \xi(y)=\sup_{y\in Y} \xi(y)<\infty
$$
and hence, by the arbitrariness of $\varphi$ and $k$, it follows that $\bar{I}<\infty$. It is straightforward to see that $\bar{I}$ is monotone, indeed if $\xi\leq \psi$, then $$\left\lbrace I(\varphi)+k:\varphi\in \mathbf{B}, k\in \mathbb{R},\ \varphi+k\leq \xi \right\rbrace\subseteq \left\lbrace I(\varphi)+k:\varphi\in \mathbf{B}, k\in \mathbb{R},\ \varphi+k\leq \psi \right\rbrace.$$ Moreover, $\bar{I}$ is translation invariant. Indeed, suppose $\xi\in B(Y,\mathcal{A})$ and $k\in \mathbb{R}$,
\begin{align*}
\bar{I}(\xi+k)&=\sup_{\varphi+a\leq \xi+k} I(\varphi)+a=\sup_{\varphi+(a-k)\leq \xi} I(\varphi)+a=\sup_{\varphi+b\leq \xi} \{I(\varphi)+b\}+k=\bar{I}(\xi)+k.
\end{align*}
Furthermore, $\bar{I}(0)=I(0)=0$ and by translation invariance it follows that $\bar{I}$ is normalized. Passing to concavity, suppose that $I$ is concave, $\xi,\psi\in B(Y,\mathcal{A})$, and $\alpha\in [0,1]$. Then,
\begin{align*}
\bar{I}( \alpha\xi+(1-\alpha)\psi)&=\sup\limits_{\varphi+k\leq \alpha\xi+(1-\alpha)\psi}\left\lbrace I(\varphi)+k\right\rbrace\\
&\geq \sup\limits_{\varphi_1+k_1\leq \xi,\ \varphi_2+k_2\leq \psi}\left\lbrace I(\alpha \varphi_1+(1-\alpha)\varphi_2)+\alpha k_1+(1-\alpha)k_2 \right\rbrace\\
& \geq \sup\limits_{\varphi_1+k_1\leq \xi,\ \varphi_2+k_2\leq \psi}\left\lbrace \alpha I(\varphi_1)+(1-\alpha)I(\varphi_2)+\alpha k_1+(1-\alpha)k_2 \right\rbrace\\
& =\alpha \sup\limits_{\varphi_1+k_1\leq \xi}\left\lbrace I(\varphi_1)+k_1 \right\rbrace+(1-\alpha) \sup\limits_{\varphi_2+k_2\leq \psi}\left\lbrace I(\varphi_2)+k_2 \right\rbrace=\alpha \bar{I}(\xi)+(1-\alpha)\bar{I}(\psi).
\end{align*}
Thus, $\bar{I}$ is concave.
\end{proof}

\subsection*{Proofs for Theorem \ref{monrecu} and sections \ref{sec:CAAA} and \ref{sec:genrect}}

We start with some notation and a lemma. For all $m\geq 0$, we denote by $\pi_m:\Omega\to \Omega$ the \textit{shift operator} defined by $\pi_m(s_1,s_2,\ldots)=(s_{m+1},s_{m+2},\ldots)$ for all $(s_1,s_2,\ldots)\in\Omega$. Moreover, for all $m\geq 1$, $\mathbf{x}=(x_0,\ldots,x_{m-1})\in X^m$, and $h\in \mathbf{H}$ we denote by $P^{\mathbf{x}}_m(h)$ the \textit{prefix consumption plan} defined as
\begin{equation}\label{prefixdefinition}
(P_m^{\mathbf{x}}(h))_t:\omega\mapsto \begin{cases}
x_t & 0\leq t <m,\\
h_{t-m}(\pi_m(\omega)) & t\geq m.
\end{cases}
\end{equation}
We have $P^{\mathbf{x}}_m(h)\in \mathbf{H}$. Indeed, if $0\leq t<m$, then $(P_m^{\mathbf{x}}(h))_t=x_t$ is constant and hence $\mathcal{G}_t$-measurable. Let $t\geq m$, clearly $h_{t-m}$ is $\mathcal{G}_{t-m}$-measurable. Now let $E=A\times S^{\infty}$ with $A\subseteq S^{t-m}$. Then, $\pi^{-1}_m(E)=S^m\times A\times S^\infty\in \mathcal{G}_t$, therefore $\pi_m$ is $(\mathcal{G}_t,\mathcal{G}_{t-m})$-measurable and hence $(P_m^{\mathbf{x}}(h))_t=h_{t-m}\circ \pi_m$ is $\mathcal{G}_t$-measurable. Therefore, $P_m^{\mathbf{x}}(h)\in \mathbf{H}$.

\begin{lemma}\label{lem_prefixes}
 Suppose $\succsim$ on $\mathbf{H}$ satisfies  axioms \hyperref[axi:weakorder]{WO}, \hyperref[axi:cont]{C}, \hyperref[axi:nontriv]{NT}, \hyperref[axi:mon]{M}, and assume that there exist a continuous $u:X\to \mathbb{R}$ and $\beta\in (0,1)$ such that the associated discounted utility $U$ represents $\succsim$ on $\mathbf{D}$. For all $h\in \mathbf{H}$, $m\geq 1$, and $\mathbf{x}\in X^m$, we have that:
\begin{enumerate}
\item For all $\omega\in \Omega$,
\begin{equation}\label{prefix_identity}
U\left(P^{\mathbf{x}}_m(h)(\omega)\right)=\sum_{\tau=0}^{m-1}\beta^\tau u(x_\tau)+\beta^m U(h(\pi_m(\omega)))
\end{equation}
\item If $\succsim$ satisfies stationarity (\hyperref[axi:stationarity]{S}), then, for all $h,g\in \mathbf{H}$,
\begin{equation}\label{equivalence_prefixes}
h\succsim g \Longleftrightarrow P_m^{\mathbf{x}}(h)\succsim P_m^{\mathbf{x}}(g).
\end{equation}
\item Suppose that $I_0:U(\mathbf{H})\to \mathbb{R}$ is a certainty equivalent such that $I_0\circ U$ represents $\succsim$, then, for all $h\in \mathbf{H}$,
\begin{equation}\label{prefix_identity2}
I_0\left(U(P^{\mathbf{x}}_m(h))\right)=\sum_{\tau=0}^{m-1}\beta^\tau u(x_\tau)+\beta^m I_0(U(h)).
\end{equation}
\end{enumerate}
\end{lemma}
\begin{proof}
Let $h\in \mathbf{H}$, $m\geq 1$, and $\mathbf{x}\in X^{m}$. We now prove \eqref{prefix_identity}.
\begin{align*}
U\left(P^{\mathbf{x}}_m(h)\right)=\sum_{t\geq 0}\beta^t u((P^{\mathbf{x}}_m(h))_t)=\sum_{t= 0}^{m-1}\beta^t u(x_t)+\sum_{t\geq m}\beta^t u(h_{t-m}\circ\pi_m)=\sum_{\tau=0}^{m-1}\beta^\tau u(x_\tau)+\beta^m U(h\circ \pi_m).
\end{align*}
To see \eqref{equivalence_prefixes}, it is sufficient to notice that $P^{\mathbf{x}}_m(h)=(x_0,(x_1,\ldots,(x_{m-1},h),\ldots)$. Then, the result follows from stationarity (\hyperref[axi:stationarity]{S}). Passing to \eqref{prefix_identity2}, let $d\in \mathbf{D}$ such that $h\sim d$. Then, by the previous point we have that $P_m^{\mathbf{x}}(h)\sim P_m^{\mathbf{x}}(d)$ and hence, by the normalization of $I_0$, the fact that $I_0\circ U$ represents $\succsim$, and  \eqref{prefix_identity},
$$
I_0(U(P_m^{\mathbf{x}}(h)))=U(P_m^{\mathbf{x}}(d))=\sum_{t= 0}^{m-1}\beta^t u(x_t)+\beta^m U(d)=\sum_{t= 0}^{m-1}\beta^t u(x_t)+\beta^m I_0(U(h)).
$$
\end{proof}

\begin{proposition}
\label{recursiveIID}
Let $\succsim$ be a binary relation on $\mathbf{H}$. If $\succsim$ satisfies axioms \hyperref[axi:weakorder]{WO}, \hyperref[axi:cont]{C}, \hyperref[axi:nontriv]{NT}, \hyperref[axi:indepdetprosp]{IDP}, \hyperref[axi:stateinddyncons]{SIDC}, \hyperref[axi:stationarity]{S}, \hyperref[axi:timesep]{TS}, and \hyperref[axi:mon]{M}, then $\succsim$ admits an IID separable recursive representation   $(V,I_{+1},u,\beta)$ with $V$ continuous in the product topology.
\end{proposition}

\begin{proof} By axioms \hyperref[axi:weakorder]{WO}, \hyperref[axi:cont]{C}, \hyperref[axi:nontriv]{NT}, \hyperref[axi:indepdetprosp]{IDP}, \hyperref[axi:stateinddyncons]{SIDC}, \hyperref[axi:stationarity]{S}, and \hyperref[axi:timesep]{TS}, standard results yield that there exists an affine, continuous, and nonconstant function $u:X\rightarrow \mathbb{R}$ and $\beta\in (0,1)$ such that $$U:d\mapsto\sum_{t\geq 0} \beta^t u(d_t),$$
represents $\succsim$ on $\mathbf{D}$. Since $u$ is continuous, cardinally unique, and $X$ is compact we can assume that $u(X)=[\beta-1,1-\beta]$, so that $U(\mathbf{D})=\left[-1,1\right]$.

Now fix $h\in \mathbf{H}$. Since $C$ is compact we have that $X=\bigtriangleup^{\sigma}(C)$ is compact as well. Therefore, given that $\succsim$ is continuous (\hyperref[axi:cont]{C}), complete and transitive (\hyperref[axi:weakorder]{WO}), we have that there must exist $x^*,x_*$ such that $x^*\succsim x \succsim x_*$ for all $x\in X$. This implies that for all $h\in \mathbf{H}$, $t\geq 0$, and $\omega\in \Omega$, we have $u(x^*)\geq u(h_t(\omega))\geq u(x_*)$ and hence,
$$
U(h_0,x^*,x^*,\ldots)\geq U(h(\omega))\geq U(h_0,x_*,x_*,\ldots)
$$
for all $\omega\in \Omega$. Therefore, by monotonicity (\hyperref[axi:mon]{M}),
\begin{equation}
(h_0,x^*,x^*,\ldots)\succsim h \succsim (h_0,x_*,x_*,\ldots)
\end{equation}
for all $h\in \mathbf{H}$. This implies that, for all $h\in \mathbf{H}$, the sets
$$\{x\in X: (h_0,x,x,\ldots)\succsim h\}\ \textnormal{and}\ \{x\in X: h\succsim (h_0,x,x,\ldots)\}$$
are not empty. Furthermore, by the continuity (\hyperref[axi:cont]{C}) of $\succsim$ they are both closed. Since $\succsim$ is a weak order, it holds, for all $h\in \mathbf{H}$,
$$\{x\in X: (h_0,x,x,\ldots)\succsim h\}\cup \{x\in X: h\succsim (h_0,x,x,\ldots)\}=X.$$
Therefore, since $X$ is connected,\footnote{This follows by observing that $X=\bigtriangleup^{\sigma}(C)$ is a convex set endowed with the weak-convergence topology, and hence it is path-connected.} we must have that
$$\{x\in X: (h_0,x,x,\ldots)\succsim h\}\cap \{x\in X: h\succsim (h_0,x,x,\ldots)\}\neq \varnothing$$
which implies that there exists $x_h \in X$ such that $(h_0,x_h,x_h,\ldots) \sim h$ for all $h\in \mathbf{H}$. Thus we can define the map
$$
V(h)=u(h_0)+\beta U\left(x_h,x_h,\ldots\right)
$$
for all $h\in \mathbf{H}$ and $V$ represents $\succsim$. Now define $I_{+1}: B(U(\mathbf{D}),S,\Sigma)\to \mathbb{R}$ as 
$$
I_{+1}(V\circ h^1)=U\left(x_h,x_h,\ldots\right). \footnote{Notice also that the following holds true
$$
B(U(\mathbf{D}),S,\Sigma)=\left\lbrace V(h^1):h\in H \right\rbrace.
$$}
$$
We verify that $I_{+1}$ is a well-defined certainty equivalent. We divide the proof into a few steps.
\begin{claim0}\label{well_defined_claim}
$I_{+1}$ is well defined.
\end{claim0}
\begin{proof}[Proof of Claim \ref{well_defined_claim}]
Let $h,g\in \mathbf{H}$ and $x_h,x_g\in X$ such that $h\sim (h_0,x_h,x_h,\ldots)$ and $g\sim (g_0,x_g,x_g,\ldots)$. First we show that for all $a,b\in \mathbb{R}$,
\begin{equation}\label{welldefined_inequality}
U(h)+a\geq U(g)+b \Longrightarrow U(h_0,x_h,x_h,\ldots)+a\geq U(g_0,x_g,x_g,\ldots)+b.
\end{equation}
Assume that $U(h)+a\geq U(g)+b$ for some $a,b\in \mathbb{R}$. For $m\geq 1$ sufficiently large we have that $\max\left\lbrace |\beta^ma|,|\beta^mb| \right\rbrace \leq 1-\beta^m$. Notice that
$$
\left\lbrace \sum_{\tau=0}^{m-1}\beta^\tau u(x_\tau):(x_0,\ldots,x_{m-1})\in X^m \right\rbrace=[\beta^m-1,1-\beta^m] 
$$
and hence, for all $i\in \left\lbrace a,b \right\rbrace$, there exists $(x^i_0,\ldots,x^i_{m-1})\in X^m$ such that $\sum_{\tau=0}^{m-1}\beta^\tau u(x^i_\tau)=\beta^m i$. Denote by $P^a_m(h)$, $P^b_m(g)$ the related prefix consumption plans as defined in \eqref{prefixdefinition}. Then, by Lemma \ref{lem_prefixes} and \eqref{prefix_identity}, we have that
$$
U(P^a_m(h))=\beta^m[a+U(h\circ \pi_m)] \ \textnormal{and}\  U(P^b_m(g))=\beta^m[b+U(g\circ \pi_m)]
$$
and hence, since $U(h)+a\geq U(g)+b$, monotonicity yields that $P^a_m(h)\succsim P^b_m(g)$. Moreover, by the second point of Lemma \ref{lem_prefixes}, we have that $P^a_m(h)\sim P^a_m(h_0,x_h,x_h,\ldots)$ and $P^b_m(g)\sim P^b_m(g_0,x_g,x_g,\ldots)$. Therefore, $P^a_m(h_0,x_h,x_h,\ldots)\succsim P^b_m(g_0,x_g,x_g,\ldots)$, and hence, by Lemma \ref{lem_prefixes} and \eqref{prefix_identity}, it follows that
$$
\beta^m[U(h_0,x_h,\ldots)+a]=U(P^a_m(h_0,x_h,\ldots))\geq U(P^b_m(g_0,x_g,\ldots))=\beta^m[U(g_0,x_g,\ldots)+b]
$$
yielding \eqref{welldefined_inequality}. Now we show that, for all $h,g\in \mathbf{H}$,
\begin{equation}\label{welldefined_implication}
\left[\forall t\geq 1,\ h_t=g_t\right] \Longrightarrow U(x_h,x_h,\ldots)=U(x_g,x_g,\ldots).
\end{equation}
Set $\delta=u(h_0)-u(g_0)$. Then, $U(h)=u(h_0)+U(g)-u(g_0)=\delta+U(g)$ and thus, by \eqref{welldefined_inequality}, we have that $U(h_0,x_h,x_h,\ldots)=\delta+U(g_0,x_g,x_g,\ldots)$ which in turn implies $U(x_h,x_h,\ldots)=U(x_g,x_g,\ldots)$ and thus the claim follows.

Now we are ready to prove that $I_{+1}$ is well defined. Suppose that $V\circ h^1=V\circ g^1$ for some $h,g\in \mathbf{H}$. Then, define $\ell\in \mathbf{H}$ as $\ell_0=h_0$ and $\ell_t=g_t$ for all $t\geq 1$. Then, by \eqref{welldefined_implication} we have that $U(x_g,x_g,\ldots)=U(x_\ell,x_\ell,\ldots)$, where $(h_0,x_\ell,x_\ell,\ldots)\sim \ell$. By definition of $\ell$, $V\circ g^1=V\circ \ell^1$. Then, it follows that $h^{s,1}\sim \ell^{s,1}$ for all $s\in S$, and hence, by stationarity (\hyperref[axi:stationarity]{S}) and $\ell_0=h_0$, we have $h^s\sim \ell^s$ for all $s\in S$. Then, by state independent dynamic consistency (\hyperref[axi:stateinddyncons]{SIDC}), $h\sim \ell$, and hence
$$
I_{+1}(V\circ h^1)=U(x_h,x_h,\ldots)=U(x_\ell,x_\ell,\ldots)=U(x_g,x_g,\ldots)=I_{+1}(V\circ g^1)
$$
and hence $I_{+1}$ is well defined.
\end{proof}
\begin{claim0}\label{claim_monotonenormalized}
$I_{+1}$ is monotone and normalized.
\end{claim0}
\begin{proof}[Proof of Claim \ref{claim_monotonenormalized}]
Let $k=U(d)$ for some $d\in \mathbf{D}$ and $x_{\circ}\in X$ such that $u(x_\circ)=0$. Then, $V((x_\circ,d))=U((x_\circ,d))=\beta k$ and hence $\beta I_{+1}(k)=\beta k$, proving that $I_{+1}$ is normalized. We prove that $I_{+1}$ is also monotone. Suppose that $\xi\geq \psi$ for some $\xi,\psi\in B(U(\mathbf{D}),S,\Sigma)$. By the same reasoning as in the proof of the previous claim $I_{+1}$ is independent of the first period $0$, thus we can assume without loss of generality that there exist $h,g\in \mathbf{H}$ such that $h_0=g_0$ and $\xi=V\circ h^1$, $\psi=V\circ g^1$. Since $V$ represents $\succsim$, we have that $h^{s,1}\succsim g^{s,1}$ for all $s\in S$. Then by \hyperref[axi:stationarity]{S} we have 
$$
h^s=(h_0,h^{s,1})\succsim (h_0,g^{s,1})=g^s
$$
for all $s\in S$, and hence, by \hyperref[axi:stateinddyncons]{SIDC}, we have that $h\succsim g$. Therefore we obtain $V(h)\geq V(g)$ which implies that $U\left(x_h,x_h,\ldots\right)\geq U(x_{g},x_{g},\ldots)$, delivering us the monotonicity of $I_{+1}$.
\end{proof}
\noindent Thus, $I_{+1}$ is a certainty equivalent. To conclude we show that $V$ is continuous in the product topology. Suppose that $h^n\to h$ in the product topology and let $\left(h^{n_m}\right)_{m\geq 0}$ be a subsequence. Then, for all $m\geq 0$, there exists $x^{n_m}\in X$ such that $h^{n_m}\sim (h_0^{n_m},x^{n_m},x^{n_m},\ldots)$. Since $X$ is compact, $\left(x^{n_m}\right)_{m\geq 0}$ admits a subsequence $\left(x^{n_{m_j}}\right)_{j\geq 0}$ converging to some $x$. For all $j\geq 0$, we have that
$$
\left(h_0^{n_{m_j}},x^{n_{m_j}},x^{n_{m_j}},\ldots\right)\sim h^{n_{m_j}}
$$
since $h^{n_{m_j}}\to h$, we have, by continuity (\hyperref[axi:cont]{C}) of $\succsim$, that $(h^0,x,x,\ldots)=\lim \left(h_0^{n_{m_j}},x^{n_{m_j}},x^{n_{m_j}},\ldots\right)\sim h$. Then, we have that
\begin{align*}
V(\lim h^{n})&=V(h)=V(h^0,x,x,\ldots)=u(h^0)+\beta U(x,x,\ldots)\\
& =\lim u\left(h_0^{n_{m_j}}\right)+\beta \lim U\left(x^{n_{m_j}},x^{n_{m_j}},\ldots\right)=\lim_{j\to \infty}V\left(h^{n_{m_j}}\right).
\end{align*}
Given the arbitrariness of $\left(h^{n_m}\right)_{m\geq 0}$, it follows that $V(h^n)\to V(h)$ as any subsequence of $\left(V(h^n)\right)_{n\geq 0}$ admits a subsequence converging to $V(h)$.
 \end{proof}

\begin{proof}[Proof of Theorem \ref{monrecu}]
\noindent $\left[(i)\Rightarrow (ii)\right]$.
By Proposition \ref{recursiveIID} we have that $\succsim$ admits an IID separable recursive representation 
$$
V:h\mapsto u(h_0)+\beta I_{+1}\left(V\circ h^1\right)
$$
for some affine, continuous, and nonconstant $u:X\to \mathbb{R}$, $\beta\in (0,1)$, certainty equivalent $I_{+1}:B(U(\mathbf{D}),S,\Sigma)\rightarrow\mathbb{R}$, and with $V$ continuous in the product topology. To ease the rest of the exposition we divide the proof in several claims.
\begin{claim}\label{I0_CE}
There exists a certainty equivalent $I_0:U(\mathbf{H})\to \mathbb{R}$ such that $I_0\circ U$ represents $\succsim$.
\end{claim}
\begin{proof}[Proof of Claim \ref{I0_CE}]
By continuity (\hyperref[axi:cont]{C}) and monotonicity (\hyperref[axi:mon]{M}), for all $h\in \mathbf{H}$ there exists $d^h\in \mathbf{D}$ such that $h\sim d^h$. This allows us to define $I_0:U(\mathbf{H})\to \mathbb{R}$ as
$$
I_0\left(U(h)\right)=U(d^h)
$$
for all $h\in \mathbf{H}$. Now we prove that $I_0$ is a well-defined certainty equivalent. By monotonicity (\hyperref[axi:mon]{M}), $I_0$ is well defined and monotone. In addition, $I_0$ is normalized. Indeed, if $k\in U(\mathbf{D})$ we have that there exists $d\in \mathbf{D}$ such that $U(d)=k$ and hence $I_0(k)=I_0(U(d))=U(d)=k$. Therefore, $I_0$ is a certainty equivalent. Moreover, $I_0$ represents $\succsim$ as $h\succsim g$ if and only if $d^h\succsim d^g$. \end{proof}
 \begin{claim}\label{I0_is_TI}
The certainty equivalent $I_0$ is translation invariant.
 \end{claim}
 Before passing to the proof we make some observations. Since $u$ is continuous, cardinally unique, and $X$ is compact we can assume that $u(X)=\left[\beta-1,1-\beta\right]$, so that $U(\mathbf{D})=\left[-1,1\right]$. We denote by $x_{\circ}$ the element of $X$ such that $u(x_{\circ})=0$. 

\begin{proof}[Proof of Claim \ref{I0_is_TI}]
First we show that for all $a,b\in\mathbb{R}$ and $\xi,\varphi\in U(\mathbf{H})$, we have
\begin{equation}\label{eq_inequality}
\xi+a\geq \varphi+b \Longrightarrow I_0(\xi)+a\geq I_0(\varphi)+b. 
\end{equation}
To this end, let $m\geq 1$ be sufficiently large so that $\max\left\lbrace |\beta^ma|,|\beta^mb|\right\rbrace\leq 1-\beta^m$. Notice that
$$
\left\lbrace \sum_{\tau=0}^{m-1}\beta^\tau u(x_\tau):(x_0,\ldots,x_{m-1})\in X^m \right\rbrace=[\beta^m-1,1-\beta^m]
$$
and hence, there exist $(x_0^a,\ldots,x_{m-1}^a),(x_0^b,\ldots,x_{m-1}^b) \in X^m$ such that
$$
\beta^m a=\sum_{\tau=0}^{m-1}\beta^\tau u(x^a_\tau)\ \textnormal{and}\ \beta^m b=\sum_{\tau=0}^{m-1}\beta^\tau u(x^b_\tau).
$$
Denote by $P^i_m(h)$ the related prefix consumption plans as defined in \eqref{prefixdefinition} for $i\in \left\lbrace a,b\right\rbrace$ and $h\in\mathbf{H}$. By Lemma \ref{lem_prefixes} and \eqref{prefix_identity2}, we have that
$$
\forall h\in \mathbf{H},\ \forall i\in \left\lbrace a,b\right\rbrace,\ I_0(U(P^i_m(h)))=\beta^mi+\beta^mI_0(U(h)).
$$
Now let $f,g\in \mathbf{H}$ such that $\xi=U(g)$ and $\varphi=U(f)$. Since $\xi+a\geq \varphi+b$, we have that $U(P^a_m(g))\geq U(P^b_m(f))$ and hence by monotonicity (\hyperref[axi:mon]{M}), it follows that
$$
\beta^m [I_0(\xi)+a]=I_0(U(P_a^m(g)))\geq I_0(U(P_b^m(f)))=\beta^m[I_0(\varphi)+b] 
$$
and hence \eqref{eq_inequality} holds. Applying Lemma \ref{extension} with $U(\mathbf{H})=\mathbf{B}$ which is convex by the affinity of $u$ and $K=U(\mathbf{D})\subseteq U(\mathbf{H})$, it follows that $I_0$ admits a monotone, normalized, and translation invariant extension $\bar{I}:B(\Omega,\mathcal{G})\to \mathbb{R}$. Therefore, it follows that $I_0$ is translation invariant.
\end{proof}
Now we prove that $I_0$ and $I_{+1}$ must satisfy condition \eqref{generalrectangularity}. 
\begin{claim}\label{Gen_rect_proof}
For all $h\in \mathbf{H}$ we have
\begin{equation}\label{GR_claim}
I_0\left(\sum_{t\geq 1}\beta^t u(h_t)\right)=\beta I_{+1}\left(s\mapsto \frac{1}{\beta}I_{0}\left(\sum_{t\geq 1}\beta^t u(h^s_t)\right)\right).
\end{equation}
\end{claim}
\begin{proof}[Proof of Claim \ref{Gen_rect_proof}]
We denote by $U_0(\mathbf{H})$ the set of all $U(h)$ with $u(h_0)=0$. It is without loss of generality to consider $U(h)\in U_0(\mathbf{H})$. For all $s\in S$, let $d^s\in \mathbf{D}$ such that $d^s\sim h^{s,1}$. By stationarity we have that $(x_\circ,h^{s,1})\sim (x_\circ,d^s)$ for all $s\in S$. Moreover, by the representations we have that
\begin{align*}
I_0(U(x_\circ, h^{s,1}))=U((x_\circ, d^{s}))=\beta U(d^s)=\beta V(h^{s,1})
\end{align*}
for all $s\in S$. Therefore, by the recursive representation and $u(h_0)=0$ we have that
$$
I_0(U(h))=U(d_h)=V(d_h)=V(h)=\beta I_{+1}\left(V\circ h^1\right)=\beta I_{+1}\left(s\mapsto \frac{1}{\beta}I_0(U(x_\circ, h^{s,1}))\right)
$$
and hence the claim follows.
\end{proof}

Finally, we prove that translation invariance of $I_{+1}$ is implied by \eqref{GR_claim}.
\begin{claim}\label{I+1transinv_proof}
The certainty equivalent $I_{+1}$ is translation invariant.
\end{claim}
\begin{proof}[Proof of Claim \ref{I+1transinv_proof}]
Let $\xi\in B(U(\mathbf{D}),S,\Sigma)$ and $k\in \mathbb{R}$ such that $\xi+k\in  B(U(\mathbf{D}),S,\Sigma)$. Now notice that $(1-\beta)\xi(s), (1-\beta)(\xi(s)+k)\in u(X)$ for all $s\in S$. Therefore, for all $s\in S$, there exist $x_s,y_s\in X$ such that $(1-\beta)\xi(s)=u(x_s)$ and $(1-\beta)(\xi(s)+k)=u(y_s)$. Define $h,g\in \mathbf{H}$ with $h_0=g_0=x_\circ$ and for all $t\geq 1$ and $(s_1,s_2,\ldots)\in\Omega$,
$$
h_t(s_1,s_2,\ldots)=x_{s_1}\ \textnormal{and}\ g_t(s_1,s_2,\ldots)=y_{s_1}.
$$
Then, we have that for all $\omega=(s_1,s_2,\ldots)\in \Omega$,
$$
\sum_{t\geq 1}\beta^t u(h_t(\omega))=\sum_{t\geq 1}\beta^t u(x_{s_1})=\frac{\beta}{1-\beta} u(x_{s_1})=\beta \xi(s_1)
$$
and, similarly,
$$
\sum_{t\geq 1}\beta^t u(g_t(\omega))=\beta [\xi(s_1)+k].
$$
Therefore, we have that for all $\omega\in \Omega$,
$$
\sum_{t\geq 1}\beta^t u(h_t(\omega))+\beta k=\sum_{t\geq 1}\beta^t u(g_t(\omega)).
$$
Now notice that for all $t\geq 1$ and $s\in S$, we have that $h_t^{s}=x_s$ and $g_t^{s}=y_s$, and hence,
$$
\sum_{t\geq 1}\beta^t u(h^{s}_t)=\beta \xi(s)\ \textnormal{and}\  \sum_{t\geq 1}\beta^t u(g^{s}_t)=\beta [\xi(s)+k].
$$
This, in turn, implies that
$$
\left(s\mapsto I_0\left(\sum_{t\geq 1}\beta^t u(h^{s}_t)\right)\right)=\beta \xi\ \textnormal{and}\ \left(s\mapsto I_0\left(\sum_{t\geq 1}\beta^t u(g^{s}_t)\right)\right)=\beta [\xi+k].
$$
\noindent Then, by generalized rectangularity and normalization of $I_0$, it follows
$$
I_0\left(\sum_{t\geq 1}\beta^t u(h_t)\right)=\beta I_{+1}\left(s\mapsto \frac{1}{\beta}I_0\left(\sum_{t\geq 1}\beta^t u(h^s_t)\right)\right)=\beta I_{+1}(\xi).
$$
Analogously, we have 
$$
I_0\left(\sum_{t\geq 1}\beta^t u(g_t)\right)=\beta I_{+1}(\xi+k).
$$
Then, by the translation invariance of $I_0$, it follows that 
$$
\beta I_{+1}(\xi)+\beta k=I_0\left(\sum_{t\geq 1}\beta^t u(h_t)\right)+\beta k=I_0\left(\sum_{t\geq 1}\beta^t u(g_t)\right)=\beta I_{+1}(\xi+k)
$$
and hence $I_{+1}$ is translation invariant.
\end{proof}

\noindent $\left[(ii)\Rightarrow (i)\right]$ Clearly, by the representations, $\succsim$ satisfies \hyperref[axi:weakorder]{WO}, \hyperref[axi:nontriv]{NT}, \hyperref[axi:indepdetprosp]{IDP}, \hyperref[axi:stationarity]{S}, and \hyperref[axi:timesep]{TS}. Moreover, by monotonicity of $I_0$, $\succsim$ satisfies  \hyperref[axi:mon]{M}, while the monotonicity of $I_{+1}$ yields that $\succsim$ satisfies \hyperref[axi:stateinddyncons]{SIDC}. To conclude, $\succsim$ satisfies \hyperref[axi:cont]{C} by the continuity in the product topology of $V$. 
\end{proof}

\begin{proof}[Proof of Corollary \ref{monrecucor}]
By Theorem \ref{monrecu}, $\succsim$ admits an ex-ante representation with a translation invariant certainty equivalent $I_0$. Then, by the affinity of $u$, we have that $U(\mathbf{H})$ is convex and for all $x\in X$, $\alpha\in [0,1]$, and $h\in \mathbf{H}$,
$$
I_0\left(\alpha \sum_{t\geq 0}\beta^tu(h_t)+(1-\alpha)\frac{u(x)}{1-\beta}\right)=I_0\left(\alpha \sum_{t\geq 0}\beta^tu(h_t)\right)+(1-\alpha)\frac{u(x)}{1-\beta}
$$
and hence, for all $y\in X$,
$$
\alpha h+(1-\alpha)x\succsim \alpha y+(1-\alpha)x\Longleftrightarrow I_0\left(\alpha \sum_{t\geq 0}\beta^tu(h_t)\right)\geq \alpha\frac{u(y)}{1-\beta}.
$$
Therefore, $\succsim$ exhibits constant absolute ambiguity aversion.
\end{proof}

\begin{proof}[Proof of Corollary \ref{variationalcor}]
By Theorem \ref{monrecu}, there exist an affine function $u:X\to\mathbb{R}$ and
translation invariant certainty equivalents $I_0$ and $I_{+1}$  such that
$\succsim$ admits an IID separable recursive representation
$(V,I_{+1},u,\beta)$ and $\succsim$ is represented by the functional
$$
h \mapsto I_0\left(\sum_{t\geq 0}\beta^t\,u(h_t)\right)
$$  
Notice that since $u$ is affine, $U(\mathbf{H})$ is convex. Observe that by \eqref{uncertaintyaversion} $I_0$ is quasiconcave. Now we show that by translation invariance $I_0$ must also be concave. To this end let $\alpha\in [0,1]$, $\xi,\psi\in U(\mathbf{H})$ and in particular $\xi=U(h)$ and $\psi=U(g)$. Now let $a=I_0(\xi)$ and $b=I_0(\psi)$, then there exists $m\geq 1$ sufficiently large such that $\max\left\lbrace |\beta^ma|,|\beta^mb|\right\rbrace \leq 1-\beta^m$. Then, there exist $(x^a_0,\ldots,x^a_{m-1}), (x^b_0,\ldots,x^b_{m-1})\in X^m$, such that
$$
\forall i\in \{a,b\},\ \sum_{\tau=0}^{m-1}\beta^\tau u(x^i_\tau)=-\beta^m i.
$$
Denote by $P^i_m(f)$ the related prefix consumption plans as defined in \eqref{prefixdefinition} for $i\in \left\lbrace a,b\right\rbrace$ and $f\in\left\lbrace h,g\right\rbrace$. By Lemma \ref{lem_prefixes} and \eqref{prefix_identity2}, we have that
$$
\forall f\in \{h,g\},\ \forall i\in \{a,b\},\ I_0(U(P^i_m(f)))=-\beta^m i+\beta^m I_0(U(f)).
$$
This implies that
$$
-\beta^m a+\beta^m I_0(U(h))=0=-\beta^m b+\beta^m I_0(U(g)).
$$
and hence, by affinity of $u$ and quasiconcavity, $I_0\left(\alpha U(P^a_m(h))+(1-\alpha)U(P^b_m(g))\right)\geq 0$
and hence, by the affinity of $u$ and \eqref{prefix_identity2},
$$
\beta^m [I_0(\alpha \xi+(1-\alpha)\psi)-\alpha a-(1-\alpha)b]\geq 0
$$
proving that $I_0$ is concave. Now we prove that also $I_{+1}$ must also be concave. Let $\xi,\xi'\in  B(U(\mathbf{D}),S,\Sigma)$ and $\alpha\in (0,1)$ and $\varphi,\psi\in U(\mathbf{H})$ be such that $\varphi(s_1,s_2,\ldots)=\beta\xi(s_1)$ and $\psi(s_1,s_2,\ldots)=\beta\xi'(s_1)$ for all $(s_1,s_2,\ldots)\in \Omega$. By generalized rectangularity (recall also the proof of Claim \ref{I+1transinv_proof}), we have that
$$
I_{0}(\varphi)=\beta I_{+1}(\xi),\ I_{0}(\psi)=\beta I_{+1}(\xi'),\ \textnormal{and}\ I_{0}(\alpha\varphi+(1-\alpha)\psi)=\beta I_{+1}(\alpha\xi+(1-\alpha)\xi')
$$
and hence, by the concavity of $I_0$, it follows that $I_{+1}(\alpha\xi+(1-\alpha)\xi')\geq \alpha I_{+1}(\xi)+(1-\alpha)I_{+1}(\xi')$ proving the concavity of $I_{+1}$.

Now applying Lemma \ref{extension} with $U(\mathbf{H})=\mathbf{B}$ and $K=U(\mathbf{D})\subseteq U(\mathbf{H})$, it follows that $I_0$ admits a concave, monotone, normalized, and translation invariant extension $\bar{I}:B(\Omega,\mathcal{G})\to \mathbb{R}$. Thus, by the results in \cite{maccheroni2006ambiguity} and \cite{cerreia2014niv} we obtain the desired variational representations. In particular, there exist cost functions $c_0:\bigtriangleup(\Omega)\to \left[0,\infty\right]$ and $c_{+1}:\bigtriangleup(S)\to \left[0,\infty\right]$ such that
$$
I_0=\min\limits_{P\in \bigtriangleup(\Omega)}\left\lbrace \mathbb{E}_{P}\left[\cdot\right]+c_0(P)\right\rbrace\ \textnormal{and}\ I_{+1}=\min\limits_{\ell\in \bigtriangleup(S)}\left\lbrace \mathbb{E}_{\ell}\left[\cdot\right]+c_{+1}(\ell)\right\rbrace,
$$
as desired.
\end{proof}

\begin{lemma}\label{lem:I1CE}
Suppose $I_0:U(\mathbf{H})\to\mathbb{R}$ is a translation invariant certainty equivalent and that there exists a functional $I_{+1}:B\!\left(U(\mathbf{D}),S,\Sigma\right)\to\mathbb{R}$ such that
$$
I_0\left(\sum_{t\geq 1}\beta^tu(h_t)\right)=\beta I_{+1}\left(s\mapsto \frac{1}{\beta}I_0\left(\sum_{t\geq 1}\beta^tu(h^s_t)\right)\right)
$$
for all $h\in \mathbf{H}$. Then $I_{+1}$ is also a certainty equivalent.
\end{lemma}
\begin{proof}
First we show that $I_{+1}$ is normalized. Since $I_0$ is translation invariant and $I_{+1}$ solves the generalized rectangularity equation, then, as proved above, $I_{+1}$ is also translation invariant. Hence it suffices to check that $I_{+1}(0)=0$. By generalized rectangularity and the normalization of $I_0$,
$$
0=I_0\left(\sum_{t\geq 1}\beta^t u(x_\circ)\right)=\beta I_{+1}\left(s\mapsto \frac{1}{\beta}I_0\left(\sum_{t\geq 1}\beta^t u(x_\circ)\right)\right)=\beta I_{+1}(0)
$$
and hence $I_{+1}$ is normalized.
\par\medskip
Let $\xi,\xi'\in  B(U(\mathbf{D}),S,\Sigma)$ with $\xi\geq \xi'$ and $\varphi,\psi\in U(\mathbf{H})$ be such that $\varphi(s_1,s_2,\ldots)=\beta\xi(s_1)$ and $\psi(s_1,s_2,\ldots)=\beta\xi'(s_1)$ for all $(s_1,s_2,\ldots)\in \Omega$. By generalized rectangularity (recall also the proof of Claim \ref{I+1transinv_proof}), we have that $I_{0}(\varphi)=\beta I_{+1}(\xi),\ I_{0}(\psi)=\beta I_{+1}(\xi')$. Since $\varphi\geq \psi$ and $I_0$ is monotone, it follows that $I_{+1}(\xi)\geq I_{+1}(\xi')$ proving the monotonicity of $I_{+1}$.
\end{proof}

\begin{proof}[Proof of Proposition \ref{lem:genrectyieldsrecu}]
By Lemma \ref{lem:I1CE}, $I_{+1}$ is monotone. For $h,g\in \mathbf{H}$, suppose that $h_0=g_0$ and $h^s\succsim g^s$ for all $s\in S$. Then, we have that,
\begin{align*}
u(h_0)+I_0\left(\sum_{t\geq 1}\beta^t u(h^{s}_t)\right)=I_0(U(h^s))\geq I_0(U(g^s))=u(g_0)+I_0\left(\sum_{t\geq 1}\beta^t u(g^{s}_t)\right)
\end{align*}
for all $s\in S$. Then, by generalized rectangularity and monotonicity of $I_{+1}$, it follows that
\begin{align*}
I_0\left(\sum_{t\geq 1}\beta^t u(h_t)\right)=\beta I_{+1}\left(s\mapsto \frac{1}{\beta}I_0\left(\sum_{t\geq 1}\beta^t u(h^{s}_t)\right)\right)\geq \beta I_{+1}\left(s\mapsto\frac{1}{\beta}I_0\left(\sum_{t\geq 1}\beta^t u(g^{s}_t)\right)\right)=I_0\left(\sum_{t\geq 1}\beta^t u(g_t)\right)
\end{align*}
and hence by translation invariance $I_0(U(h))\geq I_0(U(g))$. Thus, $h\succsim g$ and hence $\succsim$ satisfies \hyperref[axi:stateinddyncons]{SIDC}.
\end{proof}

\subsection*{Proof of Proposition \ref{monrecutgeneral2}}

Before going to the main proposition we need some additional lemmas and a proposition. For all $d\in \mathbf{D}$ and $t\geq 0$, define $d^t:=(d_t,d_{t+1},\ldots)$.
\begin{lemma}\label{zerozerozero}
If $(\succsim_{s^t})_{s^t\in H}$ is a collection of preorders that satisfies \hyperref[axi:condpref]{CP} and $d,d'\in \mathbf{D}$, then, for all $s^t\in H$, $d^t \succsim_0 d'^t$ if and only if $d\succsim_{s^t}d'.$
\end{lemma}
\begin{proof}
By \hyperref[axi:condpref]{CP} we have that
\begin{align*}
d^t\succsim_0 d'^t\Leftrightarrow (d_0,d_1,\ldots,d_{t-1},d^t)\succsim_{s^t} (d_0,d_1,\ldots,d_{t-1},d'^t)
\end{align*}
and, again by \hyperref[axi:condpref]{CP}, $(d_0,d_1,\ldots,d_{t-1},d'^t)\sim_{s^t}d'.$
Therefore, by transitivity, $d^t \succsim_0 d'^t$ if and only if $d\succsim_{s^t}d'.$
\end{proof}

Before proving Proposition \ref{monrecutgeneral2}, we state a more general version of Lemma \ref{lem_prefixes}. The proof is omitted as it is totally analogous to the one of its preceding version.

Fix $s^t\in H$, $m\geq 1$,
$\mathbf{x}=(x_0,\ldots,x_{m-1})\in X^m$, and
$h\in\mathbf{H}$. The conditional prefix plan
$P_{\mathbf{x}}^{m,s^t}(h)\in\mathbf{H}$ is defined as follows on the continuations of $s^t$,
$$
\bigl(P_{\mathbf{x}}^{m,s^t}(h)\bigr)_\tau(s^t,s'_{t+1},s'_{t+2},\ldots)
=
\begin{cases}
x_{\tau-t},
&
\text{if }t\leq\tau<t+m,\\
h_{\tau-m}
\bigl(s^t,s'_{t+m+1},\ldots\bigr),
&
\text{if }\tau\geq t+m.
\end{cases}
$$
As currently defined $P_{\mathbf{x}}^{m,s^t}(h)$ is not an actual consumption plan in $\mathbf{H}$. Indeed, it should be defined for all $\omega\in \Omega$ and it should specify its values at $\tau=0,\ldots,t-1$. Nonetheless, in what follows, we always assume that $(\succsim_{s^t})_{s^t\in H}$ satisfies \hyperref[axi:condpref]{CP}. As a consequence, any two consumption plans $f,g\in \mathbf{H}$ that extend $P_{\mathbf{x}}^{m,s^t}(h)$ would be indifferent with respect to $\succsim_{s^t}$. Therefore, in what follows, we adopt a small abuse of notation considering $P_{\mathbf{x}}^{m,s^t}(h)$ as a consumption plan, formally whenever it appears it should be read as any extension of $P_{\mathbf{x}}^{m,s^t}(h)$ that makes it a consumption plan.

\begin{lemma}\label{lem_prefixes_general}
Suppose that, for all $s^t\in H$, the preference relation
$\succsim_{s^t}$ satisfies \hyperref[axi:weakorder]{WO}, \hyperref[axi:cont]{C}, \hyperref[axi:nontriv]{NT}, \hyperref[axi:indepdetprosp]{IDP}, \hyperref[axi:stationarity]{S}, \hyperref[axi:timesep]{TS}, and \hyperref[axi:mon]{M} and $(\succsim_{s^t})_{s^t\in H}$ satisfies \hyperref[axi:condpref]{CP} and \hyperref[axi:dynacons]{DC}.
Then there exist an affine, continuous, and nonconstant function
$u: X\to\mathbb R$ and a discount factor $\beta\in(0,1)$ such that $U_t$ represents $\succsim_{s^t}$ on $\mathbf{D}$. For all $s^t\in H$, $m\geq 1$, and $\mathbf{x}=(x_0,\ldots,x_{m-1})\in X^m$,
\begin{enumerate}
\item For all $h\in\mathbf H$,
\begin{align*}
U_t\left(P_{\mathbf{x}}^{m,s^t}(h)\mid s^t\right)=\sum_{\tau=0}^{m-1}\beta^\tau u(x_\tau)+\beta^m \left(U_t(h\mid s^t)\circ \pi_m\right).
\end{align*}
\item For every $h,g\in\mathbf H$,
\begin{equation*}
    h\succsim_{s^t}g
    \quad\Longleftrightarrow\quad
    P_{\mathbf{x}}^{m,s^t}(h)
    \succsim_{s^t}
    P_{\mathbf{x}}^{m,s^t}(g).
\end{equation*}
\item Suppose that $I_{s^t}:U_t(\mathbf{H}|s^t)\to \mathbb{R}$ is a certainty equivalent such that $h\mapsto I_{s^t}\left(U_t(h\mid s^t)\right)$ represents $\succsim_{s^t}$. Then,
\begin{align*}
I_{s^t}\left(U_t\left(P_{\mathbf{x}}^{m,s^t}(h)\mid s^t\right)\right)=\sum_{\tau=0}^{m-1}\beta^\tau u(x_\tau)+\beta^m I_{s^t}\left(U_t(h\mid s^t)\right).
\end{align*}
\end{enumerate}
\end{lemma}

\begin{proposition}
\label{recursive3}
Let $(\succsim_{s^t})_{s^t\in H}$ be a collection of binary relations on $\mathbf{H}$. If, for all $s^t\in H$, $\succsim_{s^t}$ satisfies \hyperref[axi:weakorder]{WO}, \hyperref[axi:cont]{C}, \hyperref[axi:nontriv]{NT}, \hyperref[axi:indepdetprosp]{IDP}, \hyperref[axi:stationarity]{S}, \hyperref[axi:timesep]{TS}, and \hyperref[axi:mon]{M} and $(\succsim_{s^t})_{s^t\in H}$ satisfies \hyperref[axi:condpref]{CP} and \hyperref[axi:dynacons]{DC}, then $(\succsim_{s^t})_{s^t\in H}$ admits a  separable recursive representation $\left(V_{s^{t}},I_{+1,s^{t}},u,\beta\right)_{s^{t}\in H}$ with each $V_{s^t}$ continuous in the product topology.
\end{proposition}
\begin{proof} If $t=0$, then we write $\succsim_0=\succsim_{s^0}$. Since $\succsim_0$ satisfies \hyperref[axi:weakorder]{WO}, \hyperref[axi:cont]{C}, \hyperref[axi:nontriv]{NT}, \hyperref[axi:indepdetprosp]{IDP}, \hyperref[axi:stationarity]{S}, and \hyperref[axi:timesep]{TS} there exists an affine, continuous, and nonconstant function $u:X\rightarrow \mathbb{R}$ and $\beta\in (0,1)$ such that $$U_0:d\mapsto\sum_{\tau\geq 0} \beta^{\tau} u(d_\tau),$$
represents $\succsim_{0}$ on $\mathbf{D}$. By Lemma \ref{zerozerozero}, it follows that
$$
d\succsim_{s^t}d' \Longleftrightarrow \sum_{\tau\geq t} \beta^{\tau-t} u(d_\tau)\geq \sum_{\tau\geq t} \beta^{\tau-t} u(d'_\tau).
$$
To simplify the notation, for all $t\geq 0$, we define the map $U_t:\mathbf{D}\to \mathbb{R}$ as
$$
U_t:d\mapsto \sum_{\tau\geq t} \beta^{\tau-t} u(d_\tau).
$$
Since each $\succsim_{s^t}$ satisfies \hyperref[axi:mon]{M} it follows that
\begin{equation}\label{eq:stm}
\left[\forall \tau\geq t,\ \forall \omega\in \Omega,\ h_\tau(\omega)\succsim_{s^t}g_\tau(\omega)\right]\Longrightarrow h\succsim_{s^t}g.
\end{equation}
Indeed, suppose that $h_\tau(\omega)\succsim_{s^t}g_\tau(\omega)$ for some $h,g\in \mathbf{H}$ and all $\tau\geq t$ and $\omega\in \Omega$. Then,
$$
\forall \omega\in \Omega,\ \sum_{\tau\geq t}\beta^{\tau-t}u(h_\tau(\omega))\geq \sum_{\tau\geq t}\beta^{\tau-t}u(g_\tau(\omega))
$$
and hence, for all $\omega\in \Omega$, by \hyperref[axi:condpref]{CP}, it follows that $h(\omega)\succsim_{s^t}g(\omega)$ and hence, by \hyperref[axi:mon]{M}, we have that $h\succsim_{s^t}g$.
\par\medskip
Since $\succsim_0$ is a continuous complete preorder and $X$ is compact, it follows that there exist $x^*,x_*\in X$ such that $x^*\succsim_0 x\succsim_0 x_*$ for all $x\in X$. Then, it follows that $x^*\succsim_{s^t} x\succsim_{s^t} x_*$ for all $x\in X$ and $s^t\in H$. By \eqref{eq:stm}, we have that $x^*\succsim_{s^t} h \succsim_{s^t} x_*$ for all $h\in \mathbf{H}$ and $s^t\in H$. Combining this with \hyperref[axi:condpref]{CP}, we get
$$
\forall s^t\in H,\ \forall h\in \mathbf{H},\ (h_0,\ldots,h_t(s^t),x^*,x^*,\ldots)\succsim_{s^t} h\succsim_{s^t}(h_0,\ldots,h_t(s^t),x_*,x_*,\ldots).\footnote{Notice that, since each $h_t$ is $\mathcal{G}_t$-measurable, we have that $h_t(s^t,s_{t+1},\ldots)$ depends only on $s^t$, therefore, with a little abuse of notation we write $h_t(s^t)$ in place of $h_t(s^t,s_{t+1},\ldots)$.}
$$
Therefore, the sets
$$\{x\in X:(h_0,\ldots,h_t(s^t),x,x,\ldots)\succsim_{s^t} h\}\ \textnormal{and}\ \{x\in X: h\succsim_{s^t}(h_0,\ldots,h_t(s^t),x,x,\ldots)\},$$
are closed, not empty, and, moreover, given that $\succsim_{s^t}$ is a weak order, their union equals $X$. Since $X$ is connected,  we must have that
$$\{x\in X:(h_0,\ldots,h_t(s^t),x,x,\ldots)\succsim_{s^t} h\}\cap \{x\in X: h\succsim_{s^t}(h_0,\ldots,h_t(s^t),x,x,\ldots)\}\neq \varnothing$$
and hence there exists $x_h^{s^t}\in X$ such that $h\sim_{s^t} (h_0,\ldots,h_t(s^t),x_h^{s^t},x_h^{s^t},\ldots)$. Thus, for all $s^t\in H$ the map $V_{s^t}:\mathbf{H}\to \mathbb{R}$ defined by
$$
V_{s^t}:h\mapsto u(h_t(s^t))+\beta U_{t}\left(x_h^{s^t},x_h^{s^t},\ldots\right)
$$  
represents $\succsim_{s^t}$. Now we define the family of one-step-ahead certainty equivalents. In particular, for all $s^t$, define $I_{+1,s^t}: B(U_t(\mathbf{D}),S,\Sigma)\to \mathbb{R}$ as 
$$
I_{+1,s^t}\left(V_{\left(s^t,\cdot\right)}\left(h\right)\right)=U_{t}\left(x_h^{s^t},x_h^{s^t},\ldots\right).\footnote{Observe that 
$$
B(U_t(\mathbf{D}),S,\Sigma)=\left\lbrace V_{(s^t,\cdot)}(h):h\in \mathbf{H} \right\rbrace.
$$}
$$
We now prove that $I_{+1,s^t}$ is a well-defined certainty equivalent. 

\begin{claim3}\label{nonIID_welldefined}
$I_{+1,s^t}$ is well defined.
\end{claim3}
\begin{proof}[Proof of Claim \ref{nonIID_welldefined}]

We first establish the following implication. For all
$h,g\in\mathbf H$ and $a,b\in\mathbb R$,
\begin{equation}
U_t(h\mid s^t)+a\geq U_t(g\mid s^t)+b
\Longrightarrow V_{s^t}(h)+a\geq V_{s^t}(g)+b.
\label{eq:conditional-comparison}
\end{equation}
Choose $m\geq 1$ sufficiently large that $
\max\left\{|\beta^m a|,|\beta^m b|\right\}
\leq 1-\beta^m.$
Under the normalization $u(X)=[\beta-1,1-\beta]$, we have
$$
\left\{
\sum_{j=0}^{m-1}\beta^j u(x_j):
(x_0,\ldots,x_{m-1})\in X^m
\right\}
=
[\beta^m-1,1-\beta^m].
$$
Consequently, there exist $\mathbf{x}^a=(x_0^a,\ldots,x_{m-1}^a)$ and $\mathbf{x}^b=(x_0^b,\ldots,x_{m-1}^b)$ in $X^m$ such that
$$
\sum_{\tau=0}^{m-1}\beta^\tau u(x_\tau^a)=\beta^m a
\quad\text{and}\quad
\sum_{\tau=0}^{m-1}\beta^\tau u(x_\tau^b)=\beta^m b.
$$
By point 1 of Lemma \ref{lem_prefixes_general},
$$
U_t\left(P_{\mathbf{x}^a}^{m,s^t}(h)\mid s^t\right)
=
\beta^m
\left[
a+\left(U_t(h\mid s^t)\circ\pi_m\right)
\right]
\ \textnormal{and}\ 
U_t\left(P_{\mathbf{x}^b}^{m,s^t}(g)\mid s^t\right)
=
\beta^m
\left[
b+\left(U_t(g\mid s^t)\circ\pi_m\right)
\right].
$$
Hence, the hypothesis in \eqref{eq:conditional-comparison} implies $U_t\left(P_{\mathbf{x}^a}^{m,s^t}(h)\mid s^t\right)
\geq
U_t\left(P_{\mathbf{x}^b}^{m,s^t}(g)\mid s^t\right).$ Therefore, by \hyperref[axi:condpref]{CP} and \hyperref[axi:mon]{M}, we have $P_{\mathbf{x}^a}^{m,s^t}(h)
\succsim_{s^t}
P_{\mathbf{x}^b}^{m,s^t}(g).$ Let $d_h^{s^t},d_g^{s^t}\in\mathbf D$ be such that $h\sim_{s^t}d_h^{s^t}$ and
$g\sim_{s^t}d_g^{s^t}$. By point 2 of Lemma \ref{lem_prefixes_general}, we have $P_{\mathbf{x}^a}^{m,s^t}(h)
\sim_{s^t}
P_{\mathbf{x}^a}^{m,s^t}(d_h^{s^t})$
and $P_{\mathbf{x}^b}^{m,s^t}(g)
\sim_{s^t}
P_{\mathbf{x}^b}^{m,s^t}(d_g^{s^t}).$ Then, by choosing deterministic extensions of
the two prefixed deterministic plans,
\begin{align*}
\beta^m\left[V_{s^t}(h)+a\right]=
U_t\left(
P_{\mathbf{x}^a}^{m,s^t}(d_h^{s^t})
\right)\geq
U_t\left(
P_{\mathbf{x}^b}^{m,s^t}(d_g^{s^t})
\right)=
\beta^m\left[V_{s^t}(g)+b\right].
\end{align*}
Since $\beta^m>0$, implication
\eqref{eq:conditional-comparison} follows. 
\par\medskip
We now can prove that $I_{+1,s^t}$ is well defined. First, suppose that $h,g\in\mathbf H$
satisfy $h_\tau(s^t,\cdot)=g_\tau(s^t,\cdot)$ for all $\tau>t$. Then,
It follows that
$$
U_t(h\mid s^t)-u(h_t(s^t))
=
U_t(g\mid s^t)-u(g_t(s^t)).
$$
Applying \eqref{eq:conditional-comparison} yields $V_{s^t}(h)-u(h_t(s^t))
=
V_{s^t}(g)-u(g_t(s^t)).$ Thus, by the definition of $V_{s^t}$,
$$
\beta U_t(x_h^{s^t},x_h^{s^t},\ldots)=V_{s^t}(h)
-
u(h_t(s^t))=V_{s^t}(g)-u(g_t(s^t))=\beta U_t(x_g^{s^t},x_g^{s^t},\ldots).
$$
Therefore, $U_t(x_h^{s^t},x_h^{s^t},\ldots)
=
U_t(x_g^{s^t},x_g^{s^t},\ldots).$
\par\medskip
To conclude, let $h,g\in \mathbf{H}$ with $V_{(s^t,s)}(h)=V_{(s^t,s)}(g)$ for all $s\in S$. Construct $\ell\in\mathbf H$ such that $
\ell_t(s^t)=h_t(s^t)$
and $\ell_\tau(s^t,\cdot)=g_\tau(s^t,\cdot)$ for all $\tau>t$, and define $\ell$ arbitrarily elsewhere, subject to measurability. Since $g$ and
$\ell$ have the same continuation following $s^t$, the previous remark yields
\begin{equation}
U_t(x_g^{s^t},x_g^{s^t},\ldots)
=
U_t(x_\ell^{s^t},x_\ell^{s^t},\ldots).
\label{eq:h-ell-equivalents}
\end{equation}
Moreover, by \hyperref[axi:condpref]{CP}, $V_{(s^t,s)}(\ell)
=
V_{(s^t,s)}(g)
=
V_{(s^t,s)}(h)$ for all $s\in S$. Since $\ell_t(s^t)=h_t(s^t)$, \hyperref[axi:dynacons]{DC} implies $
\ell\sim_{s^t}h.$
Thus, $
V_{s^t}(\ell)=V_{s^t}(h).$ Since $\ell_t(s^t)=h_t(s^t)$, it follows that
\begin{equation}
U_t(x_\ell^{s^t},x_\ell^{s^t},\ldots)
=
U_t(x_h^{s^t},x_h^{s^t},\ldots).
\label{eq:ell-g-equivalents}
\end{equation}
Combining \eqref{eq:h-ell-equivalents} and
\eqref{eq:ell-g-equivalents}, we obtain
$$
I_{+1,s^t}\left(V_{(s^t,\cdot)}(h)\right)=U_t(x_h^{s^t},x_h^{s^t},\ldots)
=
U_t(x_g^{s^t},x_g^{s^t},\ldots)=I_{+1,s^t}\left(V_{(s^t,\cdot)}(g)\right),
$$
and hence $I_{+1,s^t}$ is well defined.
\end{proof}

\begin{claim3}\label{nonIID_monotone}
$I_{+1,s^t}$ is monotone and normalized.
\end{claim3}
\begin{proof}[Proof of Claim \ref{nonIID_monotone}]
First, we prove that each $I_{+1,s^t}$ is normalized. Let $d\in \mathbf{D}$, then 
$$
\frac{u(x^*)}{1-\beta}\geq U_t(d)\geq \frac{u(x_*)}{1-\beta}.
$$
By continuity of $\succsim_{s^t}$, there exists $x_d\in X$ such that $U_t(d)=u(x_d)/(1-\beta)=U_t(x_d,x_d,\ldots)$. Therefore, $d\sim_{s^t}(x_d,x_d,\ldots)$, and hence 
$$
I_{+1,s^t}(U_t(d))=U_t(x_d,x_d,\ldots)=U_t(d).
$$
Now we prove that $I_{+1,s^t}$ is monotone. Suppose that $V_{(s^t,s)}(h)\geq V_{(s^t,s)}(g)$ for all $s\in S$. Let $\ell\in \mathbf{H}$ be such that $\ell_t(s^t)=g_t(s^t)$ and $\ell_\tau(s^t,\ldots)=h_\tau(s^t,\ldots)$ for all $\tau>t$. By the same steps as in the proof of Claim \ref{nonIID_welldefined}, we have that
\begin{equation}\label{equation_monI1st}
U_t\left(x^{s^t}_h,x^{s^t}_h,\ldots\right)= U_t\left(x^{s^t}_\ell,x^{s^t}_\ell,\ldots\right).
\end{equation}
By \hyperref[axi:condpref]{CP}, for all $s\in S$, we have $V_{(s^t,s)}(\ell)=V_{(s^t,s)}(h)\geq V_{(s^t,s)}(g)$. Thus, by \hyperref[axi:dynacons]{DC}, we have that $V_{s^t}(\ell)\geq V_{s^t}(g)$ and hence $\beta I_{+1,s^t}(V_{(s^t,\cdot)}(\ell))\geq \beta I_{+1,s^t}(V_{(s^t,\cdot)}(g))$. By \eqref{equation_monI1st}, it follows, 
$$
I_{+1,s^t}(V_{(s^t,\cdot)}(h))=U_t\left(x^{s^t}_h,x^{s^t}_h,\ldots\right)=U_t\left(x^{s^t}_\ell,x^{s^t}_\ell,\ldots\right)=I_{+1,s^t}(V_{(s^t,\cdot)}(\ell))\geq I_{+1,s^t}(V_{(s^t,\cdot)}(g))
$$
proving the claim.
\end{proof}
Let $s^t\in H$, $x\in X$, and $h\in \mathbf{H}$, we show that $V((x,h)_{s^t})=u(x)+\beta V_{s^t}(h)$.\footnote{Notice $(x,h)_{s^t}=P_x^{1,s^t}(h)$.} Let $d^{s^t}_h\in \mathbf{D}$ be such that $h\sim_{s^t}d^{s^t}_h$. Then, by \hyperref[axi:stationarity]{S}, \hyperref[axi:condpref]{CP}, point 2 of Lemma \ref{lem_prefixes_general}, $(x,h)_{s^t}\sim_{s^t}(x,d^{s^t}_h)_{s^t}$ and hence
$$
V((x,h)_{s^t})=V((x,d^{s^t}_h)_{s^t})=U_t((x,d^{s^t}_h)_{s^t})=u(x)+\beta U_t(d^{s^t}_h)=u(x)+\beta V_{s^t}(h) 
$$
where $(x,d^{s^t}_h)_{s^t}$ is an extension of the prefix taken to be in $\mathbf{D}$. To conclude we show that $V_{s^t}$ is continuous in the product topology. Suppose that $h^n\to h$ in the product topology and let $\left(h^{n_m}\right)_{m\geq 0}$ be a subsequence. Then, for all $m\geq 0$, there exists $x^{n_m}\in X$ such that $h^{n_m}\sim_{s^t} (h_0^{n_m},\ldots,h_t^{n_m}(s^t),x^{n_m},x^{n_m},\ldots)$. Since $X$ is compact, $\left(x^{n_m}\right)_{m\geq 0}$ admits a subsequence $\left(x^{n_{m_j}}\right)_{j\geq 0}$ converging to some $x$. For all $j\geq 0$, we have that
$$
\left(h_0^{n_{m_j}},\ldots,h_t^{n_{m_j}}(s^t),x^{n_{m_j}},x^{n_{m_j}},\ldots\right)\sim_{s^t} h^{n_{m_j}}
$$
since $h^{n_{m_j}}\to h$, we have, by continuity of $\succsim_{s^t}$, that $$(h_0,\ldots,h_t(s^t),x,x,\ldots)=\lim (h_0^{n_{m_j}},\ldots,h_t^{n_{m_j}}(s^t),x^{n_{m_j}},x^{n_{m_j}},\ldots)\sim_{s^t} h.$$ Then, we have that
\begin{align*}
V_{s^t}(h)&=V_{s^t}(h_0,\ldots,h_t(s^t),x,x,\ldots)=u(h_t(s^t))+\beta U_t(x,x,\ldots)\\
& =\lim_{j\to \infty} u\left(h_t^{n_{m_j}}(s^t)\right)+\beta \lim_{j\to \infty} U_t\left(x^{n_{m_j}},x^{n_{m_j}},\ldots\right)=\lim_{j\to \infty}V_{s^t}\left(h^{n_{m_j}}\right).
\end{align*}
Given the arbitrariness of $\left(h^{n_m}\right)_{m\geq 0}$, it follows that $V_{s^t}(h^n)\to V_{s^t}(h)$ as any subsequence of $\left(V_{s^t}(h^n)\right)_{n\geq 0}$ admits a subsequence converging to $V_{s^t}(h)$.
\end{proof}

\begin{proof}[Proof of Proposition \ref{monrecutgeneral2}]
\noindent $\left[(i)\Rightarrow (ii)\right]$. By Proposition \ref{recursive3}, it follows that $(\succsim_{s^t})_{s^t\in H}$ admits a separable recursive representation $\left(V_{s^t},I_{+1,s^t},u,\beta\right)_{s^t\in H}$.

\begin{claim2}\label{Ist_CE}
For all $s^t\in H$, there exists a certainty equivalent $I_{s^t}:U_t(\mathbf{H}|s^t)\to \mathbb{R}$ such that
$$h\mapsto I_{s^t}\left(\sum_{\tau\geq t} \beta^{\tau-t} u(h_\tau(s^t,\cdot))\right)$$ 
represents $\succsim_{s^t}$.
\end{claim2}
\begin{proof}[Proof of Claim \ref{Ist_CE}]
Fix $s^t\in H$. By continuity, monotonicity, and \hyperref[axi:condpref]{CP} for all $h\in H$ there exists $d^{s^t}_h\in \mathbf{D}$ such that $h\sim_{s^t}d^{s^t}_h$. We define $I_{s^t}:U_t(\mathbf{H}|s^t)\to \mathbb{R}$ such that
$$
\forall h\in \mathbf{H},\ I_{s^t}\left(U_t(h|s^t)\right)=U_t(d^{s^t}_h).
$$
By following the same steps as in the proof of Claim \ref{I0_CE} of Theorem \ref{monrecu} it is immediate to see that $I_{s^t}$ is monotone and normalized, and so a certainty equivalent.
\end{proof}

\begin{claim2}\label{Ist_transsupnorm}
For all $s^t\in H$, the certainty equivalent $I_{s^t}:U_t(\mathbf{H}|s^t)\to \mathbb{R}$ is translation invariant.
\end{claim2}
\begin{proof}[Proof of Claim \ref{Ist_transsupnorm}]
Following the same proof of Claim \ref{I0_is_TI} of Theorem \ref{monrecu}, by interchanging $I_0$ with $I_{s^t}$ and employing Lemma \ref{lem_prefixes_general}, we have that for all $h,g\in \mathbf{H}$ and $a,b\in \mathbb{R}$, 
\begin{equation}\label{inequality_generalcase}
U_t(h|s^t)+a\geq U_t(g|s^t)+b\Longrightarrow I_{s^t}\left(U_t(h|s^t)\right)+a\geq I_{s^t}\left(U_t(g|s^t)\right)+b.
\end{equation}
Thus, by Lemma \ref{extension}, the claim follows.
\end{proof}
\begin{claim2}\label{claimgenrectgen}
For all $s^t\in H$ and $h\in \mathbf{H}$,
\begin{equation}\label{genrectanggen}
I_{s^t}\left(\sum_{m\geq 1} \beta^{m} u(h_{t+m}(s^t,\cdot))\right)=\beta I_{+1,s^t}\left(s\mapsto\frac{1}{\beta}I_{ (s^{t},s)}\left(\sum_{m\geq 1} \beta^{m} u(h_{t+m}(s^t,s,\ldots))\right)\right).
\end{equation}
\end{claim2}
\begin{proof}[Proof of Claim \ref{claimgenrectgen}]
The proof is analogous to that of Claim \ref{Gen_rect_proof} in the proof of Theorem \ref{monrecu},
\begin{align*}
I_{s^t}\left(\sum_{m\geq 1}\beta^m u(h_{t+m}(s^t,\ldots))\right)&=I_{s^t}\left(\sum_{m\geq 0}\beta^m u(h_{t+m}(s^t,\ldots))\right)-u(h_t(s^t))=V_{s^t}(h)-u(h_t(s^t))\\
&=\beta I_{+1,s^t}(s\mapsto V_{(s^t,s)}(h))=\beta I_{+1,s^t}(s\mapsto I_{(s^t,s)}(U_{t+1}(h|(s^t,s))))
\end{align*}
where the last equality follows from
$$
V_{(s_t,s)}(h)=V_{(s_t,s)}(d^{(s^t,s)}_h)=U_{t+1}(d^{(s^t,s)}_h)=I_{(s^t,s)}(U_{t+1}(h|(s^t,s)))
$$
and hence the claim follows.
\end{proof}

\begin{claim2}\label{claim_gentransinvariance}
For all $s^t\in H$, the certainty equivalent $I_{+1,s^t}$ is translation invariant.
\end{claim2}
\begin{proof}[Proof of Claim \ref{claim_gentransinvariance}]
Finally, the proof of translation invariance of $I_{+1}$ given in
Theorem \ref{monrecu} applies at every history. Fix $s^t\in H$ and let
$\xi,\xi+k\in B(U_t(\mathbf{D}),S,\Sigma)$. Let $h,g\in\mathbf H$
such that
$$
\sum_{m\geq 1}\beta^m u(h_{t+m}(s^t,\cdot))=\beta\xi
\qquad\text{and}\qquad
\sum_{m\geq 1}\beta^m u(g_{t+m}(s^t,\cdot))=\beta(\xi+k).
$$
By generalized rectangularity and the translation invariance of $I_{s^t}$, we obtain
\begin{align*}
\beta I_{+1,s^t}(\xi+k)
&=I_{s^t}\left(\sum_{m\geq 1}\beta^m u(g_{t+m}(s^t,\cdot))\right)=I_{s^t}\left(\sum_{m\geq 1}\beta^m u(h_{t+m}(s^t,\cdot))+\beta k\right)\\
&=I_{s^t}\left(\sum_{m\geq 1}\beta^m u(h_{t+m}(s^t,\cdot))\right)+\beta k=\beta I_{+1,s^t}(\xi)+\beta k.
\end{align*}
Since $\beta>0$, it follows that $I_{+1,s^t}(\xi+k)=I_{+1,s^t}(\xi)+k$, thus $I_{+1,s^t}$ is translation invariant.
\end{proof}
\noindent $\left[(ii)\Rightarrow (i)\right]$ Given the separable recursive representation of $(\succsim_{s^t})_{s^t\in H}$, it is immediate to see that $(\succsim_{s^t})_{s^t\in H}$ satisfies \hyperref[axi:condpref]{CP}. Moreover, by monotonicity of each $I_{+1,s^t}$ it also follows that $(\succsim_{s^t})_{s^t\in H}$ satisfies \hyperref[axi:dynacons]{DC}. Stationarity follows from the stationary identity in Definition 3. The rest of the axioms follows immediately from the representations and the same steps provided in the proof of Theorem \ref{monrecu}.  \end{proof}

\bibliographystyle{apalike}
\bibliography{ref}

\end{document}